\pdfoutput=1

\documentclass[aip,reprint,graphicx]{revtex4-1}

\usepackage{graphicx}

\usepackage[utf8]{inputenc}

\draft

\begin{document}

\title{The Heidelberg Compact Electron Beam Ion Traps} 

\author{P.~Micke}
\email[Contact for correspondence: ]{peter.micke@mpi-hd.mpg.de}
\affiliation{Max-Planck-Institut f\"ur Kernphysik, Saupfercheckweg 1, 69117 Heidelberg, Germany}
\affiliation{Physikalisch-Technische Bundesanstalt, Bundesallee 100, 38116 Braunschweig, Germany}

\author{S.~K\"uhn}
\affiliation{Max-Planck-Institut f\"ur Kernphysik, Saupfercheckweg 1, 69117 Heidelberg, Germany}

\author{L.~Buchauer}
\affiliation{Max-Planck-Institut f\"ur Kernphysik, Saupfercheckweg 1, 69117 Heidelberg, Germany}

\author{J.~R.~Harries}
\affiliation{National Institutes for Quantum and Radiological Science and Technology, SPring-8, Kouto 1-1-1, Sayo-cho, Sayo-gun, Hyogo, 679-5148 Japan}

\author{T.~M.~B\"ucking}
\affiliation{Max-Planck-Institut f\"ur Kernphysik, Saupfercheckweg 1, 69117 Heidelberg, Germany}

\author{K.~Blaum}
\affiliation{Max-Planck-Institut f\"ur Kernphysik, Saupfercheckweg 1, 69117 Heidelberg, Germany}

\author{A.~Cieluch}
\affiliation{Max-Planck-Institut f\"ur Kernphysik, Saupfercheckweg 1, 69117 Heidelberg, Germany}

\author{A.~Egl}
\affiliation{Max-Planck-Institut f\"ur Kernphysik, Saupfercheckweg 1, 69117 Heidelberg, Germany}

\author{D.~Hollain}
\affiliation{Max-Planck-Institut f\"ur Kernphysik, Saupfercheckweg 1, 69117 Heidelberg, Germany}

\author{S.~Kraemer}
\affiliation{Max-Planck-Institut f\"ur Kernphysik, Saupfercheckweg 1, 69117 Heidelberg, Germany}

\author{T.~Pfeifer}
\affiliation{Max-Planck-Institut f\"ur Kernphysik, Saupfercheckweg 1, 69117 Heidelberg, Germany}

\author{P.~O.~Schmidt}
\affiliation{Physikalisch-Technische Bundesanstalt, Bundesallee 100, 38116 Braunschweig, Germany}
\affiliation{Institut f\"ur Quantenoptik, Leibniz Universit\"at Hannover, Welfengarten 1, 30167 Hannover, Germany}

\author{R.~X.~Sch\"ussler}
\affiliation{Max-Planck-Institut f\"ur Kernphysik, Saupfercheckweg 1, 69117 Heidelberg, Germany}

\author{Ch.~Schweiger}
\affiliation{Max-Planck-Institut f\"ur Kernphysik, Saupfercheckweg 1, 69117 Heidelberg, Germany}

\author{T.~Stöhlker}
\affiliation{GSI Helmholtzzentrum f\"ur Schwerionenforschung, Planckstraße 1, 64291 Darmstadt, Germany}
\affiliation{Helmholtz-Institut Jena, Fr\"obelstieg 3, 07743 Jena, Germany}
\affiliation{Institut f\"ur Optik und Quantenelektronik, Friedrich-Schiller-Universit\"at Jena, Max-Wien-Platz 1, 07743 Jena, Germany}

\author{S.~Sturm}
\affiliation{Max-Planck-Institut f\"ur Kernphysik, Saupfercheckweg 1, 69117 Heidelberg, Germany}

\author{R.~N.~Wolf}
\altaffiliation[Current address: ]{ARC Centre of Excellence for Engineered Quantum Systems, School of Physics, The University of Sydney, NSW 2006, Australia}
\affiliation{Max-Planck-Institut f\"ur Kernphysik, Saupfercheckweg 1, 69117 Heidelberg, Germany}

\author{S.~Bernitt}
\email[Contact: ]{sven.bernitt@mpi-hd.mpg.de}
\affiliation{Max-Planck-Institut f\"ur Kernphysik, Saupfercheckweg 1, 69117 Heidelberg, Germany}
\affiliation{Institut f\"ur Optik und Quantenelektronik, Friedrich-Schiller-Universit\"at Jena, Max-Wien-Platz 1, 07743 Jena, Germany}

\author{J.~R.~Crespo~L\'opez-Urrutia}
\affiliation{Max-Planck-Institut f\"ur Kernphysik, Saupfercheckweg 1, 69117 Heidelberg, Germany}

\begin{abstract}
Electron beam ion traps (EBIT) are ideal tools for both production and study of highly charged ions (HCI). In order to reduce their construction, maintenance, and operation costs we have developed a novel, compact, room-temperature design, the Heidelberg Compact EBIT (HC-EBIT). Four already commissioned devices operate at the strongest fields (up to 0.86 T) reported for such EBITs using permanent magnets, run electron beam currents up to 80\,mA and energies up to 10\,keV. They demonstrate HCI production, trapping, and extraction of pulsed Ar$^{16+}$ bunches and continuous 100 pA ion beams of highly charged Xe up to charge state ${29+}$, already with a 4\,mA, 2\,keV electron beam. Moreover, HC-EBITs offer large solid-angle ports and thus high photon count rates, e.~g., in x-ray spectroscopy of dielectronic recombination in HCIs up to Fe$^{24+}$, achieving an electron-energy resolving power of $E/\Delta E > 1500$ at 5\,keV. Besides traditional on-axis electron guns, we have also implemented a novel off-axis gun for laser, synchrotron, and free-electron laser applications, offering clear optical access along the trap axis. We report on its first operation at a synchrotron radiation facility demonstrating resonant photoexcitation of highly charged oxygen.
\end{abstract}


\maketitle

\section{Introduction}\label{I}
Highly charged ions (HCI) constitute a large class of atomic systems, since each element has as many ionization states as it has protons -- in a sense this extends the periodic table by a further dimension. As a consequence of virialization in deep gravitational potentials \cite{bykov_equilibration_2008}, HCIs are the predominant form in which most of the elements appear in the visible universe, be it in or around stars \cite{lisse_charge_2001}, galaxies and their clusters \cite{hitomi_collaboration_solar_2017}, or in the vast expanses of the intergalactic medium\cite{shull_baryon_2012, danforth_low-z_2008, nicastro_missing_2008}. Consequently, for many decades the study of HCIs has been essential for astrophysics and astronomy\cite{grotian_zur_1939,edlen_deutung_1943,edlen_spectra_1947}, and many examples of recent laboratory work continuously show its importance not only for astrophysics (see, e.~g., \cite{beiersdorfer_laboratory_2003-1,beiersdorfer_laboratory_2003,bernitt_unexpectedly_2012,shah_laboratory_2016} and references therein), but also for plasma and fusion research (e.~g., in \cite{suckewer_identification_1982,kaufman_magnetic-dipole_1983,edlen_forbidden_1984,finkenthal_forbidden_1984,beiersdorfer_high-resolution_1993,iwamae_polarization_2007,beiersdorfer_iter_2010,hoarty_observations_2013,rosen_validation_2014,beiersdorfer_highly_2015,beiersdorfer_iter_2017}). In atomic physics, HCI studies often deal with fundamental interactions due to relativistic effects, quantum electrodynamics (QED) and nuclear-size contributions which are all enhanced by several orders of magnitude \cite{shabaev_hyperfine_1994, shabaev_ground-state_1997, tupitsyn_relativistic_2003,vogel_aspects_2013} compared to neutral or singly charged systems. This, for instance, facilitated the precise determination of the electron mass \cite{sturm_high-precision_2014, kohler_electron_2015} and the electron magnetic moment \cite{kohler_isotope_2016} as well as stringent QED tests\cite{klaft_precision_1994, seelig_ground_1998,gumberidze_quantum_2005,sturm_$g$_2011,kubicek_transition_2014}. Moreover, several proposals contemplate HCIs as ideal laboratory probes of a possible variation of the fine-structure constant $\alpha$\cite{schiller_hydrogenlike_2007, berengut_enhanced_2010, berengut_electron-hole_2011, berengut_transitions_2011, berengut_optical_2012, dzuba_ion_2012, yudin_magnetic-dipole_2014, safronova_highly_2014, safronova_highly_2014-1, safronova_atomic_2014,ong_optical_2014,dzuba_optical_2015, dzuba_actinide_2015, oreshkina_hyperfine_2017} and as frequency references for optical clocks\cite{berengut_highly_2012,dzuba_high-precision_2012,dzuba_erratum:_2013,derevianko_highly_2012,ludlow_optical_2015,lopez-urrutia_frequency_2016,yu_scrutinizing_2016} superior to state-of-the-art optical lattice or singly-charged ion clocks. The very low polarizability of their electronic wave function explains their insensitivity to both spurious external perturbations and laser-induced light shifts. Suitable forbidden optical transitions have been theoretically identified, and preliminary laboratory determinations of their energies carried out \cite{windberger_identification_2015}. Other current applications are tumor ion therapy \cite{kraft_radiobiological_2003} and EUV nanolithography \cite{harilal_spectral_2006,osullivan_spectroscopy_2015, windberger_analysis_2016,torretti_optical_2017}. Nevertheless, the HCI research community has remained rather small as HCI production is perceived as demanding and, indeed, there are still open challenges before it becomes a standard routine.

One of the first ever mentions of an `ion trap' in the literature, by Pierce and others in the 1940's\cite{heil_notitle_1946,field_control_1947,pierce_theory_1954}, describes a system in which atoms were ionized by electron impact and their ions radially trapped by the negative space charge potential of the electron beam and axially by cylindrical electrodes forming an axial potential well. They carried out investigations on ionic space charge and its effect on electron-beam propagation. Donets\cite{donets_ultrahigh_1969,donets_electron_1985,donets_electron_1990} and Arianer\cite{arianer_multiply_1975,arianer_orsay_1976} introduced the electron beam ion source, with the addition of a magnetic field for electron-beam compression. The modern electron beam ion trap (EBIT) based upon this was developed in the 1980's at Lawrence Livermore National Laboratory (LLNL) by Marrs and Levine\cite{marrs_measurement_1988, levine_electron_1988, levine_use_1989}. An EBIT operates by means of a focused mono-energetic electron beam. Emitted from an electron gun, this beam is accelerated, guided along the axis of a set of cylindrical electrodes (referred to as drift tubes), decelerated, and dumped on a collector electrode. A strong axial magnetic field of increasing flux density compresses the beam to a diameter of a fraction of a millimeter at the trap center, resulting in an extremely high current density (in certain EBITs on the order of 10$^4$\,A/cm$^2$) -- the key feature for efficient ionization and subsequent ion trapping. The beam energy results from the potential difference between the cathode and the central trap electrode, with corrections due to the space charge of both electron beam and trapped ions, the work function of the materials used for the electrodes, and adjacent potentials. Neutral atoms can be injected as an atomic beam by introducing a gas or a volatile organometallic compound through a differentially pumped injection system. Other techniques are also applied, employing a laser ion source\cite{trinczek_laser_2006} or a wire-probe target\cite{elliott_wire_1995}. Ionized by electron-beam impact, the ions are immediately trapped radially by the negative space charge of the compressed electron beam itself, as well as by the magnetic field. Axially, the confinement is controlled by the electrostatic potentials applied to the drift tubes. Charge breeding is realized by further sequential ionization of the trapped ions by the electron beam. The highest possible charge state is limited by its kinetic energy. Finally, the charge-state distribution is determined by the ionization and recombination rates, depending on the electron beam energy and density, the confinement time, and the background pressure\cite{penetrante_evolution_1991}. Details on the operating principle can be found elsewhere in the literature\cite{gillaspy_highly_2001,currell_physics_2003,currell_physics_2005}.

With direct optical access to the confined ion cloud, mono-energetic excitation, narrow ion-charge distributions and small source-volume sizes, EBITs have become essential spectroscopic tools in the last three decades, from the pioneering work on x-rays at LLNL, down to the optical range, there as well as in other groups \cite{morgan_observation_1995,crespo_lopez-urrutia_direct_1996, serpa_kr_1997, crespo_lopez-urrutia_precision_1998, crespo_lopez-urrutia_nuclear_1998, utter_measurement_2000,porto_uv_2000, porto_erratum:_2006,beiersdorfer_hyperfine_2001, watanabe_magnetic_2001, draganic_high_2003,hosaka_laser_2004,orts_exploring_2006, soria_orts_zeeman_2007,lopez-urrutia_visible_2008,liang_extreme-ultraviolet_2009,beilmann_prominent_2011,beilmann_major_2013,baumann_contributions_2014, bekker_forbidden_2015,bekker_identifications_2015,lapierre_relativistic_2005,mackel_laser_2011,brenner_lifetime_2007,brenner_transition_2009,schnorr_coronium_2013,windberger_identification_2015}. Beyond electron-impact excitation and ionization of the trapped ions utilizing the electron beam, photoexcitation and photoionization by mono-energetic photons at synchrotrons \cite{simon_photoionization_2010,simon_resonant_2010,rudolph_x-ray_2013,epp_single-photon_2015,steinbrugge_absolute_2015} and free-electron laser (FEL) facilities \cite{epp_soft_2007,epp_x-ray_2010,bernitt_unexpectedly_2012} have also been reported, in part using the EBIT magnetic trapping mode\cite{beiersdorfer_magnetic_1996}, for which the electron beam is switched off. However, these experiments are limited by Doppler broadening due to high ion temperatures (10$^5$ to 10$^7$\,K). To overcome this limitation, HCIs can be extracted from an EBIT and loaded into Paul\cite{schwarz_cryogenic_2012} or Penning traps\cite{schneider_confinement_1994,gruber_evidence_2001,hobein_evaporative_2011, repp_pentatrap:_2012, brewer_capture_2013, sturm_high-precision_2017} where advanced cooling techniques can be applied. Recently, Ar$^{13+}$ ions have been re-trapped in a cryogenic radio-frequency trap\cite{schmoger_deceleration_2015} and sympathetically cooled down to below 100\,mK by laser-cooled Be$^{+}$ Coulomb crystals \cite{schmoger_coulomb_2015,schmoger_kalte_2017}. This 8-orders-of-magnitude cooling will finally allow high-precision spectroscopy on HCIs as it is routinely performed with atoms and singly charged ions, and the application of the most sensitive techniques for detection, like quantum logic spectroscopy\cite{schmidt_spectroscopy_2005}, aiming at resolving the natural linewidth of forbidden optical transitions, or direct frequency-comb spectroscopy from the optical to the extreme ultra-violet (XUV) range\cite{nauta_towards_2017}. Furthermore, extraction and subsequent detection of ions can be used to determine the charge-state distribution in the trap and has been used to investigate resonant photoionization by synchrotron radiation\cite{simon_photoionization_2010,simon_resonant_2010,steinbrugge_absolute_2015}. EBITs also operate as versatile HCI sources in experiments investigating charge transfer processes\cite{xue_kinematically_2014}, HCI-surface interactions\cite{mcdonald_observation_1992,aumayr_emission_1993, ritter_pit_2012,makgato_highly_2013,ritter_novel_2013,wilhelm_interatomic_2017}, and for rapid charge breeding of radioactive ions\cite{dilling_mass_2006,ettenauer_first_2011}.

\begin{figure}
\includegraphics[width=0.9\linewidth]{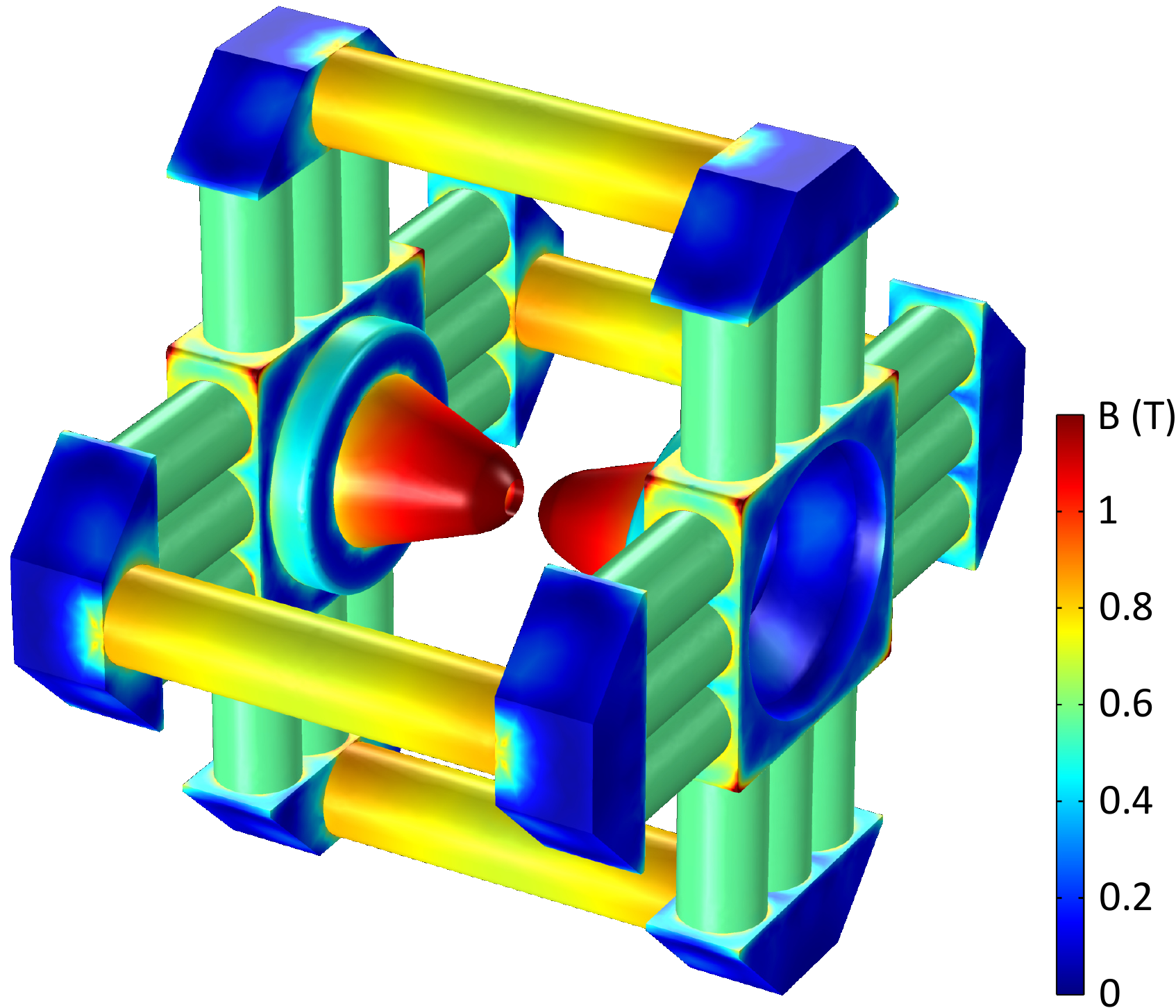}
\caption{HC-EBIT magnetic structure\label{MagSys}: Color-coded absolute magnetic flux density on the outer surfaces (simulated with COMSOL \cite{noauthor_notitle_2013}). Permanent magnets (appearing in green shades) produce the field, which is guided by magnetic-steel (blue shades) and soft-iron (yellow shades) parts, concentrated towards the gap at the trap center (reaching there 0.86\,T), and displays its maximum at the tip (red shades) of the soft-iron pole pieces.}%
\end{figure}

\begin{figure*}
\includegraphics[width=0.9\linewidth]{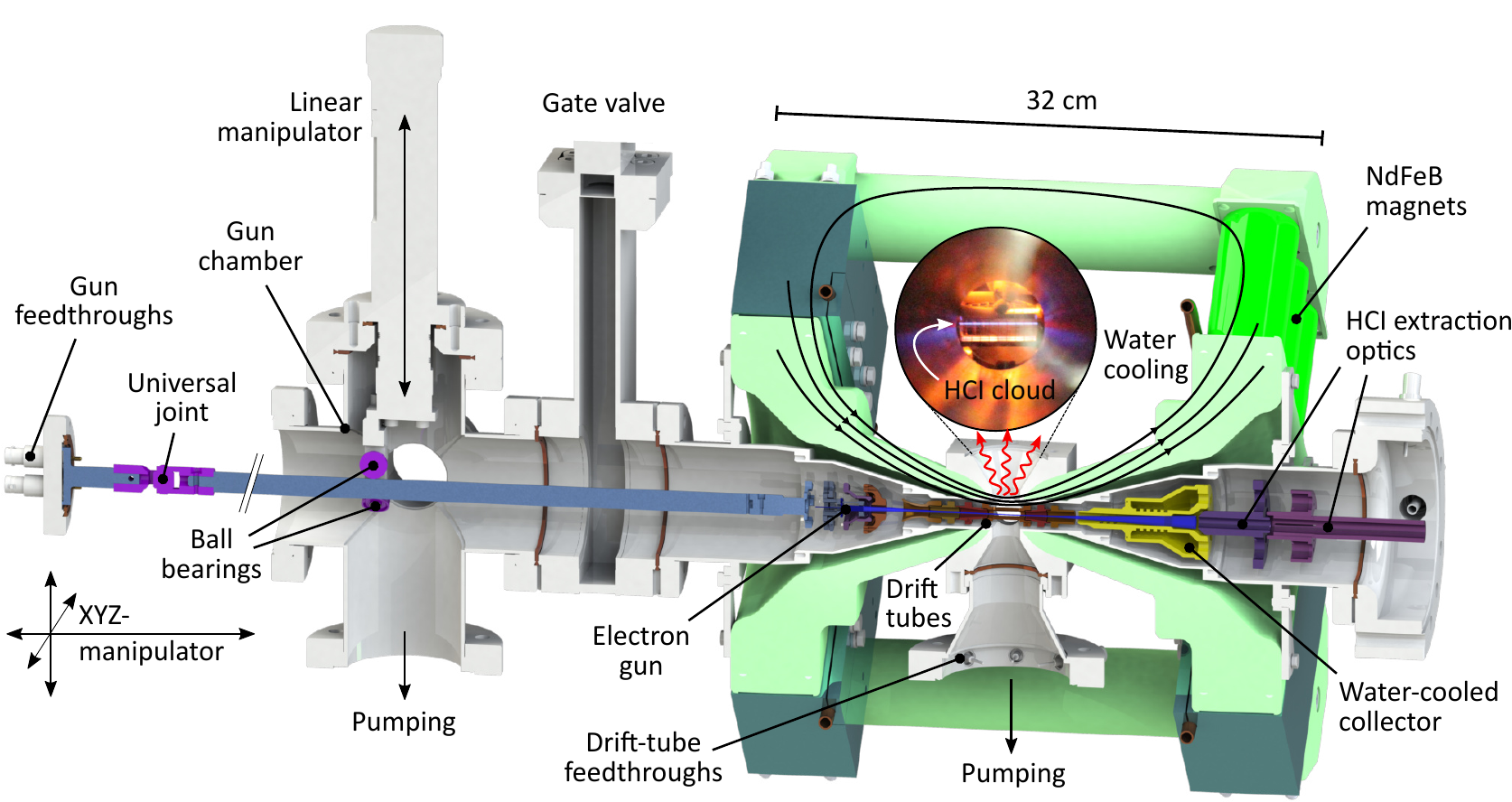}
\caption{Simplified cross section of the HC-EBIT\label{Design} design. The magnetic field is generated by 72 NdFeB magnets arranged in eight arrays of $3\times3$ magnets (dark green, encased in protecting aluminum cartridges), guided and focused by soft-iron and magnetic-steel elements (light green). In the gap around the trap center the field reaches 0.86\,T. The electron gun is mounted on a long titanium rod and positioned with an XYZ- and a linear manipulator. The drift-tube assembly and the collector are mounted inside the central vacuum chamber. The inset shows a photograph of a bluish fluorescing HCI cloud excited by electron-beam impact.}%
\end{figure*}

Most of the reported experiments have been carried out using high-performance EBITs, employing superconducting magnets with flux densities from 3 to 8\,T. Room-temperature EBITs with permanent magnets have also been developed in order to reduce size as well as costs and to ease operation. After the pioneering apparatus built in Paris by Khodja and Briand \cite{khodja_warm_1997}, soon several others followed in Dresden \cite{ovsyannikov_first_1999, zschornack_compact_2013}, Tokyo\cite{motohashi_compact_2000}, Belfast \cite{watanabe_belfast_2004}, Shanghai \cite{xiao_very_2012}, Clemson\cite{takacs_diagnostic_2015}, and at NIST \cite{hoogerheide_miniature_2015}. Related devices based on permanent magnets were also recently developed \cite{ovsyannikov_main_2016}.

Here, we report on a novel class of devices with a stronger magnetic field than for any other previously built room-temperature EBIT. The operation with the newly designed electron gun, drift-tube assembly, and collector results in excellent performance parameters. Requirements of low cost, low maintenance, reliable and stable operation, high-numerical-aperture optical access for spectroscopy, easy transportability, and compact size have been fulfilled. These points are crucial for providing HCIs to a variety of new experiments and are in part prerequisites for measurements at synchrotron-radiation sources and FELs. Following our first prototype, already serving as an HCI source for a Penning trap, we have commissioned three further devices, PTB-EBIT, PolarX-EBIT and Tip-EBIT, of the upgraded HC-EBIT design on which this report will mainly focus.

\section{Design}\label{D} 

\subsection{Magnetic structure and central vacuum chamber}

\begin{figure}
\includegraphics[width=0.8\linewidth]{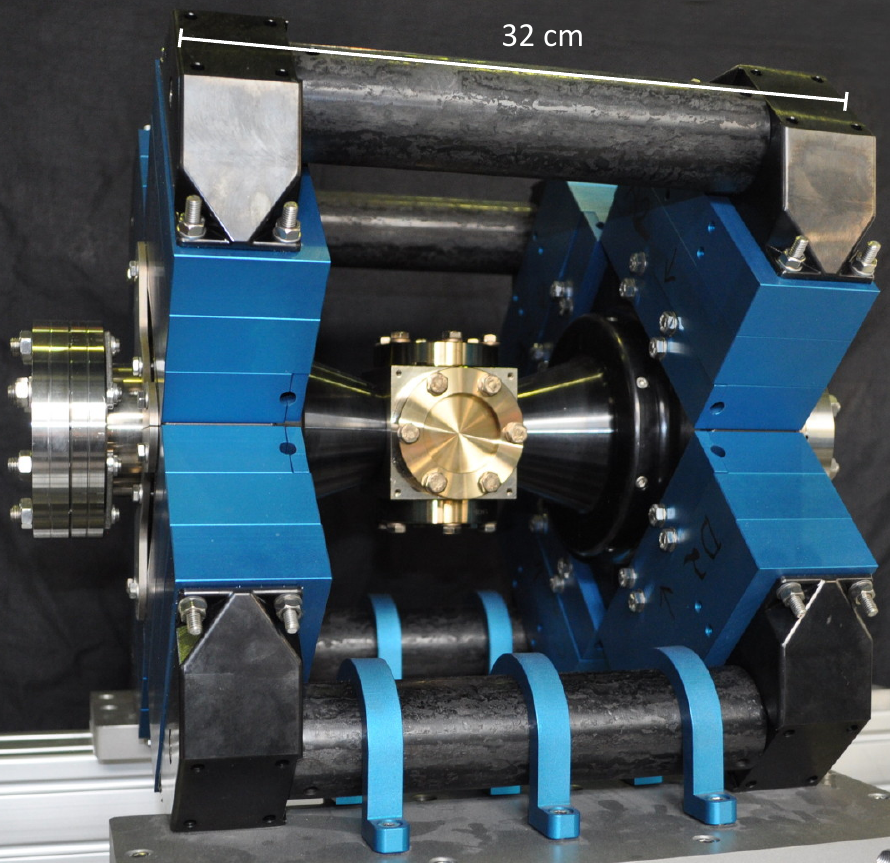}
\caption{Photograph of one of the commissioned magnetic structures with vacuum chamber\label{PTBebit}. Blue-anodized aluminum cartridges house the NdFeB magnets. Soft-iron and magnetic-steel elements were burnished to prevent rusting.}%
\end{figure}

\begin{figure*}
\includegraphics[width=1\linewidth]{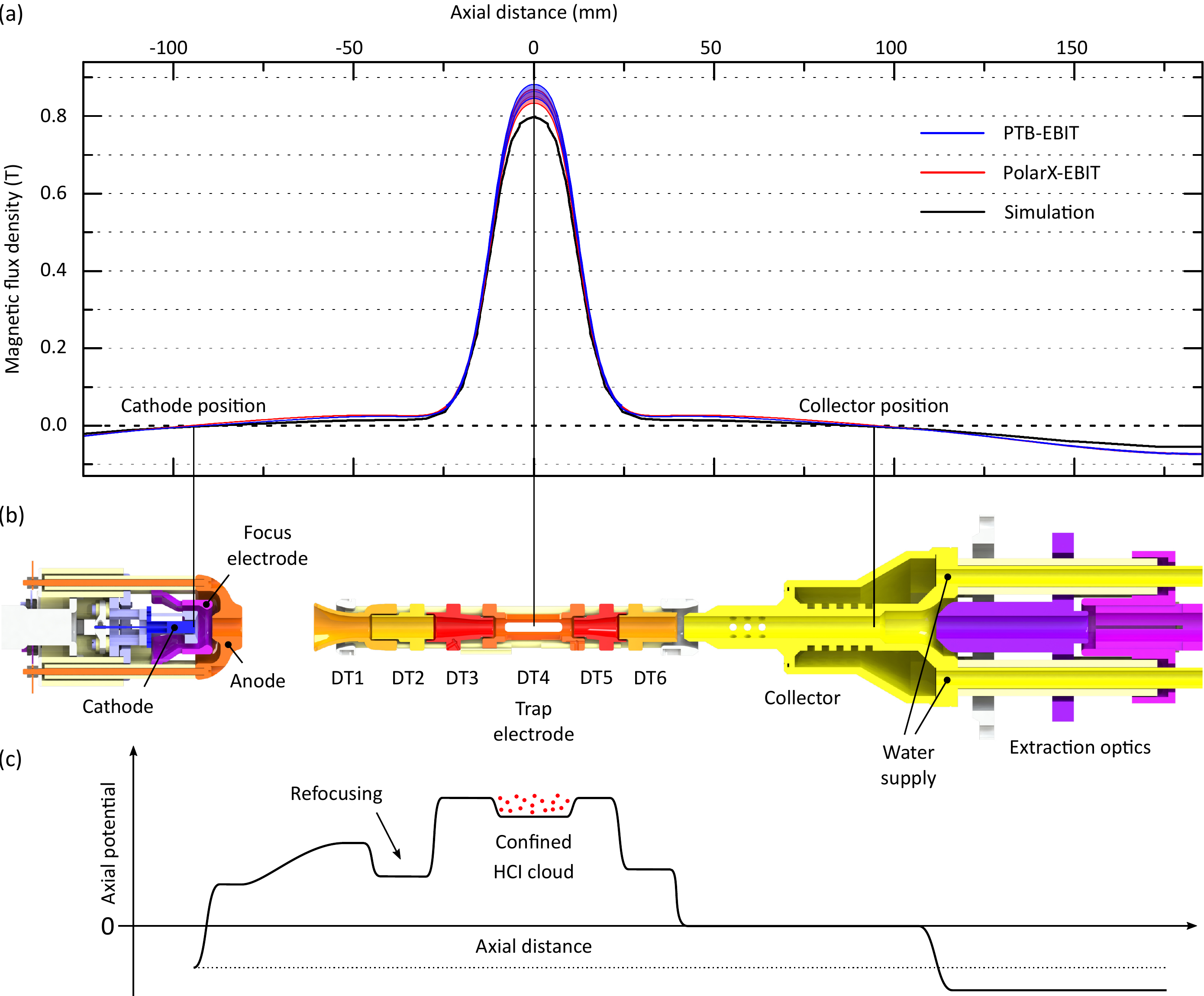}
\caption{Axial magnetic field and electrostatic system of the HC-EBIT design\label{El_sys}. (a) Simulated and measured magnetic flux densities on the electron beam axis are shown. The COMSOL simulations apparently underestimate the saturation magnetization and permeability of materials or the magnetization of the permanent magnets, resulting in slightly higher experimentally achieved values. The measurements (with gaussmeter 7010, F.W. Bell) match within uncertainties except for minor deviations between 30 and 110\,mm from the trap center. (b) The electrostatic system including on-axis Pierce-type electron gun, drift-tube assembly with drift tubes 1 to 6 (DT1 to DT6), water-cooled collector, and ion extraction optics composed of two individual tubes for focusing is displayed. (c) An illustration of a typical axial electrostatic potential curve is shown. Here, DT1 to DT3 are used to refocus the electron beam while DT3 to DT5 provide the axial ion trapping. The extraction optics electrodes are biased to a lower voltage than the cathode in order to prevent electrons from passing through.}%
\end{figure*}

We chose a magnetic structure with a discrete four-fold rotational symmetry, allowing for a compact overall size and sufficient space around the trap (see Fig.~\ref{MagSys}). Close access for detectors, spectrometers, pumps, and target injection is offered through four radial DN40CF ports (see Figs.~\ref{Design} and \ref{PTBebit}) machined on four sides of the central cubical chamber (side length of 70\,mm). Four arrays of NdFeB permanent disk magnets (dark green in Fig.~\ref{Design}) for each of the two poles generate the magnetic field. Each disk magnet (N45 quality, diameter of 45\,mm, height of 30\,mm) is magnetized along its cylinder axis. The arrays consist of three parallel stacks of three magnets each and are mounted between magnetic-steel parts connected to four flux-return rods (soft iron) and two hollow conical pole pieces (soft iron), respectively, constituting the entire yoke (light green in Fig.~\ref{Design}). These pole pieces guide and concentrate a nearly fully rotationally symmetric magnetic field into the trap region while they are close to magnetic saturation. The pieces are bisected along their symmetry axis and are mounted surrounding two conical vacuum chamber sections, which contain the electron gun and the collector. These sections are welded at their respective narrow ends to the central cube, which has an inner bore of 16\,mm along the trap axis, and houses the drift-tube assembly. The two conical sections widen to DN63CF flanges and form together with the cube a symmetric 405-mm-long chamber. The pole pieces are fitted into conical bores on either side of the cube to reduce the magnetic gap (19\,mm long, 19\,mm bore diameter), while remaining outside of the vacuum. The sharp-edged geometry at their tips helps to leak out the field efficiently. The brittle magnets are mounted in stacked aluminum cartridges (see Fig.~\ref{PTBebit}) providing mechanical protection and a water-cooling system to keep them below their Curie temperature of 80\,$^\circ$C during bake-out of the vacuum chamber.

The whole magnetic structure has a footprint of 320\,mm\,$\times$\,350\,mm with a height of 350\,mm and generates a magnetic flux density of more than 0.86\,T at the trap center. Finite-element simulations (COMSOL\cite{noauthor_notitle_2013}) were used to optimize the setup with various simultaneous requirements in mind: maximum flux density at trap center with zero field at the cathode position for electron beam compression according to Herrmann's theory \cite{herrmann_optical_1958} as well as sufficient flux density between trap center and both electron gun and collector for beam transport. The magnetic flux density at the trap center was found to be limited by the material properties of the soft-iron pole pieces, resulting in the choice of three layers of magnets on each array. Adding a fourth layer would only increase the flux by about 3\,\% according to our simulations. Fig.~\ref{El_sys}(a) compares our simulations and the measured fields, which are slightly higher due to underestimated permeability and saturation strength of both soft iron and magnetic steel as well as the magnetization of the NdFeB magnets. At 93.4(2.1)\,mm axial distance from the trap center, two locations with zero field are chosen for cathode and collector positions. At larger distances the field reverses direction and increases again in magnitude before decaying to zero (see Fig.~\ref{El_sys}(a)).

\begin{figure}
\includegraphics[width=0.9\linewidth]{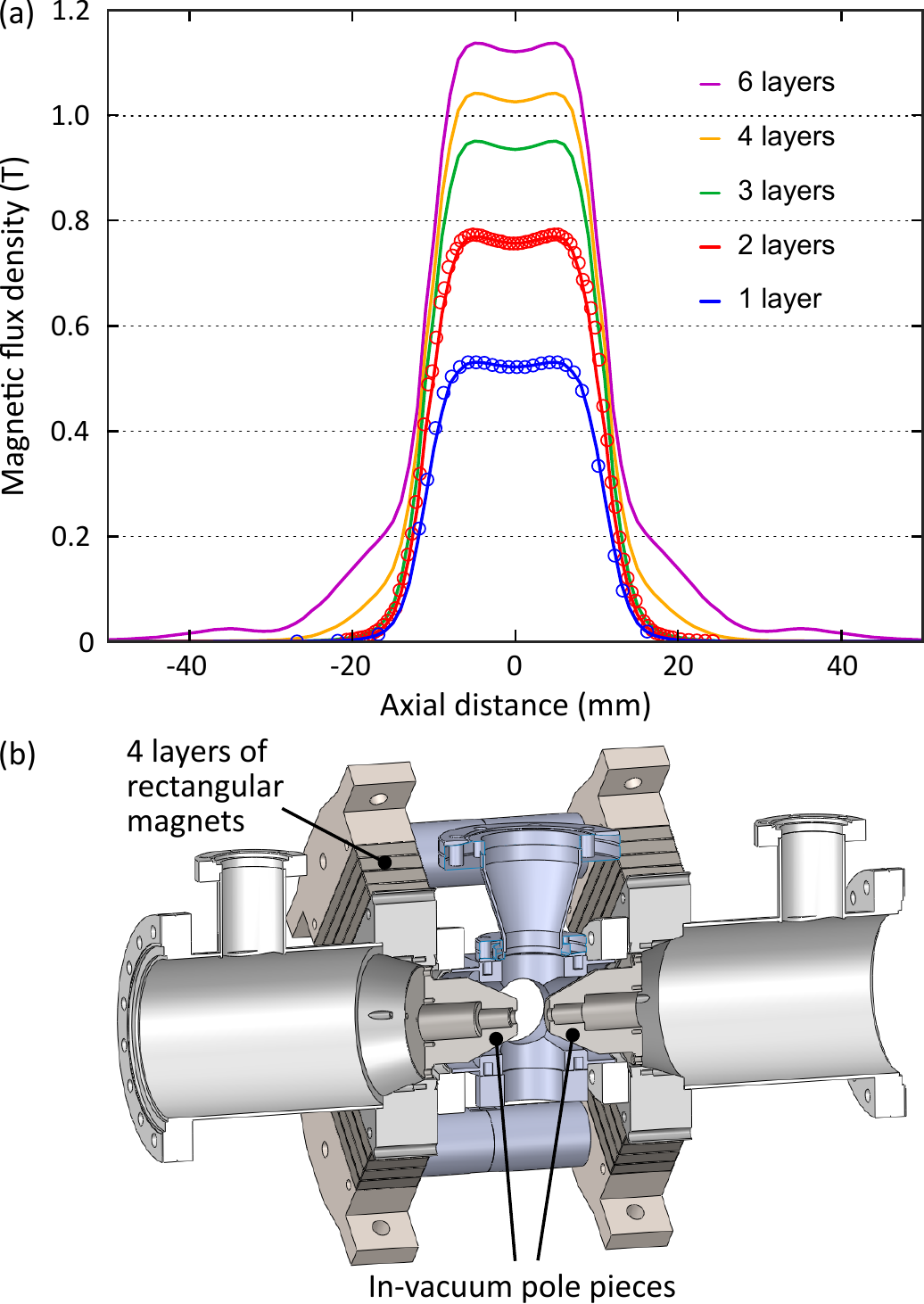}
\caption{SPring-8 prototype\label{spring-8}. (a) Axial magnetic field simulations (solid lines, FEMM) and measurements (circles) up to 2 magnet layers. The good agreement promises achievable magnetic flux densities of more than 1\,T. Differing from the HC-EBIT design, the SPring-8 prototype uses rectangular magnets and in-vacuum pole pieces allowing for a narrower bore as shown in (b).}%
\end{figure}

A further increase of the magnetic field requires in-vacuum pole pieces for a narrower bore. We follow this approach with a very similar prototype at SPring-8\cite{harries_notitle_nodate} using rectangular magnets and in-vacuum vanadium permendur pole pieces (VIC International) with a bore of only 8\,mm diameter to reach field strengths of over 1\,T (see Fig.~\ref{spring-8}). Measurements with this scalable apparatus and FEMM simulations of up to 6 magnet layers are compared in Fig.~\ref{spring-8}(a) indicating good agreement.

\subsection{Assembling the magnetic structure}
Mounting the strong permanent magnets requires particular care, since forces acting on them can suddenly appear. Their non-linear dependence on the magnetic gap easily makes them uncontrollable, causing injuries and destruction of the magnets. The magnets should only be individually unpacked and handled, and mounting of the magnetic structure should make use of tools that keep body parts far away from a potential squeezing hazard. To reduce such risks, we developed the following procedure: First, the central vacuum chamber is horizontally mounted on two holders for the DN63CF flanges which are firmly fixed to a heavy table (wooden or aluminum-top). The holders permit, when loosened, a rotation around the central axis of the vacuum chamber. Then, the bisected conical pole pieces are inserted into the grooves of the central cube and the similarly bisected square-profile yoke elements are mounted around the rim of the pole pieces with a 90\,$^\circ$ rotation between their respective cuts. The square-profile elements are screwed to the pole pieces and also to two rings welded to the vacuum chamber. The resulting stable structure protects the weld seam at the cube from a potential bending or torsion of the conical vacuum chamber extensions when, e.~g., the DN63CF flanges are tightened. Prior to the mounting of the magnets, their field strengths were measured for their proper arrangement in the magnet arrays to accomplish the rotational symmetry for the magnetic field of the EBIT. Furthermore, the polarity of each magnet was clearly marked on it. The magnets can be individually handled, provided that no magnetic parts are close nearby. For the mounting procedure, we have designed a mounting tool capable of magnetically grabbing a single magnet on a magnetic-steel piston brazed to the end of a M10 threaded rod. This piston slides inside a 0.5\,m-long hollow brass rod when screwing the threaded rod in or out. A grabbed magnet rests on the far end of the tool, since the tool has the same outer diameter of 45\,mm as the magnet. By moving the magnetic piston inwards, it slowly separates from the magnet and releases it safely. The tool can be held with two hands at a safe distance from the magnet and facilitates a careful manual insertion of each magnet into the corresponding hole of the cartridge. When the grabbed magnet approaches the yoke or the already mounted magnet beneath, eddy currents induced in the aluminum cartridges and the limited air-flow through the tight gap between the magnet and the cartridge reduce the attracting force and prevent damage when the magnet-magnet gap is closed. After partial filling of one array, repulsive forces start to appear at certain gap separations between the magnet being inserted and the layer of magnets below. The mounting tool helps to push the magnet further in to overcome this repulsion. Then, the force immediately  reverses direction and the magnet is again pulled inwards. After completing two arrays by filling all six cartridges on one side, the outer flux-return rod is installed to complete that quarter of the magnetic structure. In this last step, the soft-iron rod, preinstalled to the magnetic-steel caps on either side, is lowered to the two magnet arrays by temporarily using long threaded rods to guide this motion. Further threaded rods are screwed through threads in the magnetic-steel caps and rest on the two topmost cartridges when the outer yoke part has slid towards the arrays. By screwing these rods out, the outer yoke part approaches the arrays further and the magnetic gap can be closed in a controlled way. After that, one proceeds to fill the cartridges on the opposite side to keep magnetic forces balanced. By rotating the whole structure as described above, one can always work in horizontal direction. In principle, the stacked structure of cartridges also allows for removing possibly damaged magnets by sliding the aluminum cartridges sideways and gradually reducing the magnetic forces between the stacked magnets. However, no magnets were damaged during the assembly of all three EBITs of the new HC-EBIT design. We strongly recommend that the procedure described above is only performed by trained personnel following strict safety rules in order to avoid serious injuries.

\subsection{Electron gun}

\subsubsection{Choice of the cathode}

The strong electron beam of an EBIT, typically hundreds of mA for a superconducting EBIT, can only be reliably sustained over long periods of operation by the use of thermionic dispenser cathodes. Among many available options, barium-impregnated tungsten dispenser cathodes have proven to be the most suitable cathode type for EBITs. Due to their low work function ($\approx2$\,eV), they can operate at rather low temperatures of around 1300\,K and yield current emission densities of up to 10\,A/cm$^2$. This dispenser-type cathode has been the most widely used choice in telecommunications, radar, aerospace, and related industries for decades. A depletion of barium on the emission surface is compensated by a constant replenishment from the tungsten-matrix reservoir of the dispenser cathode. This principle affords routine lifetimes of about 100,000 hours of operation even at strong emission currents. The material itself is not brittle, can be machined and electron-beam welded. Molybdenum, alumina ceramic, and OFHC copper (provided it is not too close to the hot cathode) are compatible as adjacent materials inside an electron gun and do not induce poisoning of the cathode. Reliable commercial suppliers exist in the market. Compared to alternative cathode materials (e.g., IrCe or crystalline materials), barium-impregnated dispenser-type cathodes offer longer lifetimes, more stable emission behavior, resilience against experimental accidents, and a lower operation temperature. In particular the latter advantage eases constrains on the choice of adjacent materials in the electron gun, allows for a better residual pressure there, and allows for a higher electron beam compression in a given magnetic field according to Herrmann's optical theory\cite{herrmann_optical_1958}.

\subsubsection{Pierce-geometry on-axis electron gun}

\begin{figure}
\includegraphics[width=0.9\linewidth]{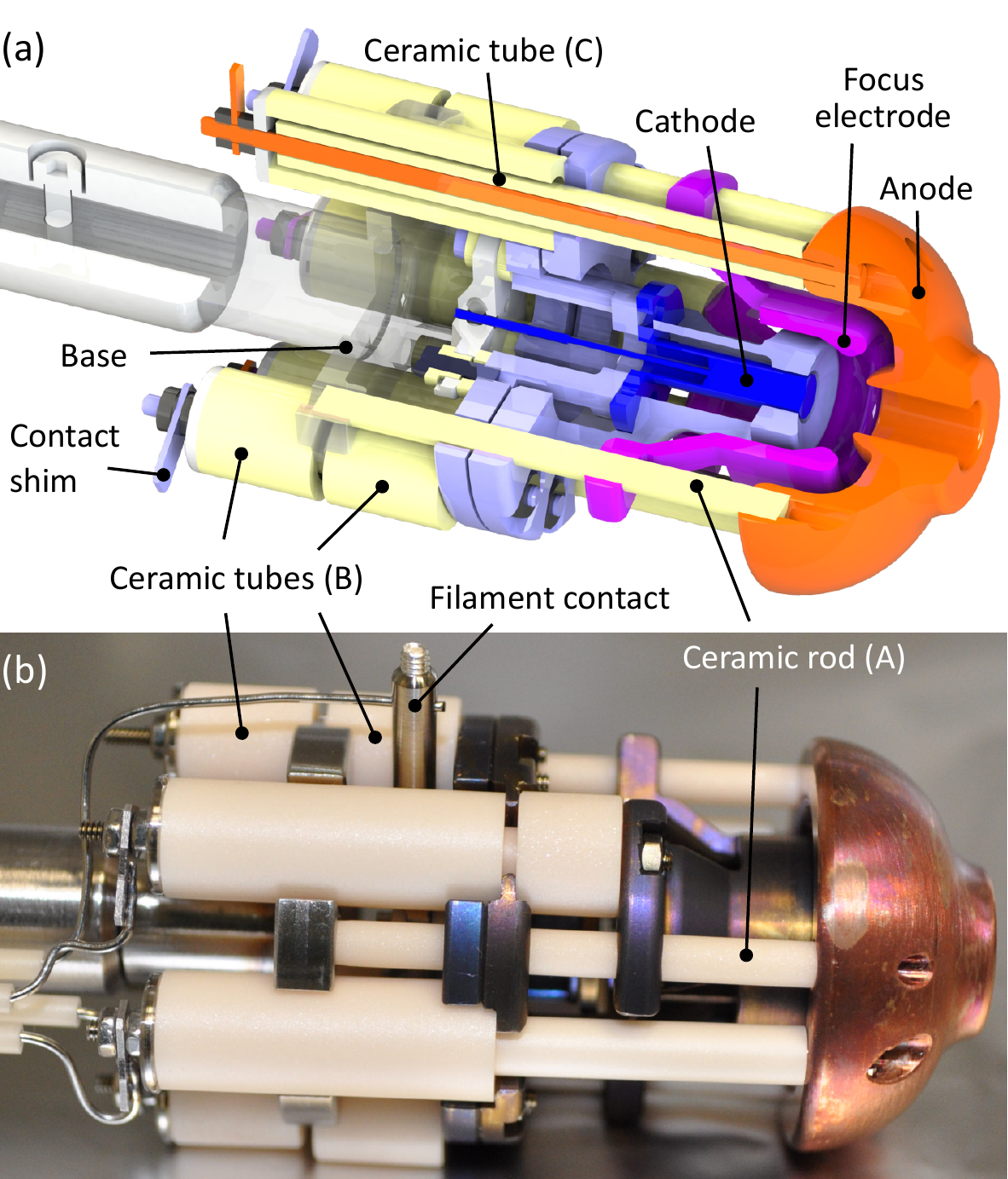}
\caption{Cross-sectional view (a) and photograph (b) of the Pierce-type on-axis electron gun\label{on-axis_gun}. The electron-gun electrodes are held by a nested alumina-ceramic structure supported by a stainless-steel base on the back (transparent in (a)), providing high-voltage insulation with sufficiently large creeping distances. See text for further details.}%
\end{figure}

With its open structure, our new on-axis electron-gun design aims at being even more sturdy against unexpected thermal loads, voltage spikes, and discharges than our earlier models. This is particularly important for the parts that are close to the cathode, since gaps have to be small. The Pierce-type\cite{pierce_rectilinear_1940} electron gun (shown in Fig.~\ref{on-axis_gun}) is, despite the rotational symmetry of the electrodes around the electron beam, rotationally asymmetric to facilitate a more compact design. It employs a 3.4 mm-diameter thermionic barium-impregnated tungsten dispenser cathode (type `M' coating) with a concave spherical radius of 8.2\,mm, which is clamped between two molybdenum parts. An OFHC copper anode, a molybdenum focus electrode, the cathode assembly, and a stainless-steel base are stacked along three alumina-ceramic rods (A in Fig.~\ref{on-axis_gun}) with small dimensional tolerances to define the centering of the electrodes. These rods also act as spacers between anode and base. The cathode assembly is tightened with two threaded rods to the base against alumina-ceramic tubes (B) which electrically insulate and set the correct distances. Thinner alumina-ceramic tubes (C) avoid spark-over to adjacent metal parts. The focus electrode and anode are similarly and independently clamped to the cathode assembly. Anode, focus electrode, and cathode are contacted on the back of the electron gun by shims on which wires are spot-welded. The cathode-heater filament is contacted with a triangular plate on the back of the cathode assembly through a lateral pin. All threaded rods, screws, and nuts near the cathode are made from molybdenum to resist the high temperature of the cathode of around 1400\,K during operation. A crucial prerequisite for a reliable and strong long-term emission current, $I_b$, is a very low local residual pressure. The chosen open gun structure allows good pumping to facilitate this. The anode, with an aperture of 5\,mm, is located at a distance of 5\,mm in front of the cathode and controls the emission current independently of the potential difference between cathode and trap-center electrode, determining the beam energy in the trap. Between cathode and anode, the focus electrode compensates potential distortions to improve the beam transport. Furthermore, the focus-electrode voltage is dynamically adjusted to stabilize the emitted electron-beam current to better than $ \Delta I_b / I_b \approx 10^{-4}$ on a $> 100$\,ms time scale to suppress long-term drifts when performing hour-long measurements. At the beginning of a measurement series, the gun position is carefully adjusted to optimize the current, current density, and beam transmission. For this, the electron gun is mounted on the far end of a horizontal titanium rod attached to an XYZ-manipulator by a universal joint on a DN40CF flange, also holding the high-voltage (HV) feedthroughs for the gun. The rod angle is set by a vertical, linear manipulator which supports the rod with a pivot point located roughly at the middle of the rod. Moreover, the gun can be completely retracted, sliding on ball bearings on that pivot point, into a dedicated gun chamber, which can be separated from the main chamber with a gate valve. In this setup, the on-axis electron gun has generated more than 80\,mA of electron-beam current without approaching the temperature limit of the thermionic cathode.

\subsubsection{Off-axis electron gun}
In the last decade, photoexcitation and photoionization in EBITs have become reliable techniques for investigating HCIs. 
The energy resolution in such studies allows to resolve the natural line widths of the transitions involved and to study asymmetric line profiles due to quantum interference. 
Monoenergetic x-ray photon beams from both free-electron lasers \cite{epp_soft_2007,epp_x-ray_2010,bernitt_unexpectedly_2012} and synchrotron radiation sources \cite{simon_photoionization_2010,simon_resonant_2010,rudolph_x-ray_2013,epp_single-photon_2015,steinbrugge_absolute_2015} have been used for this purpose. Optical laser spectroscopy inside an EBIT has also been demonstrated \cite{mackel_laser_2011,schnorr_coronium_2013}.
Usually, the preferred method for these types of experiments is introducing the photon beam through the collector along the trap axis, since a maximum overlap of the photon beam and the trapped HCI cloud is desirable. As a consequence, the photon beam is dumped onto the electron gun and, accordingly, discarded. Further disadvantages are vacuum degradation and HV problems, since the photon beam can excessively produce photoelectrons at some of the electron-gun electrodes. Sputtered ions and pressure increases may damage the cathode and reduce its lifetime. 
Furthermore, alignment can be difficult.

\begin{figure}
	\includegraphics[width=0.9\linewidth]{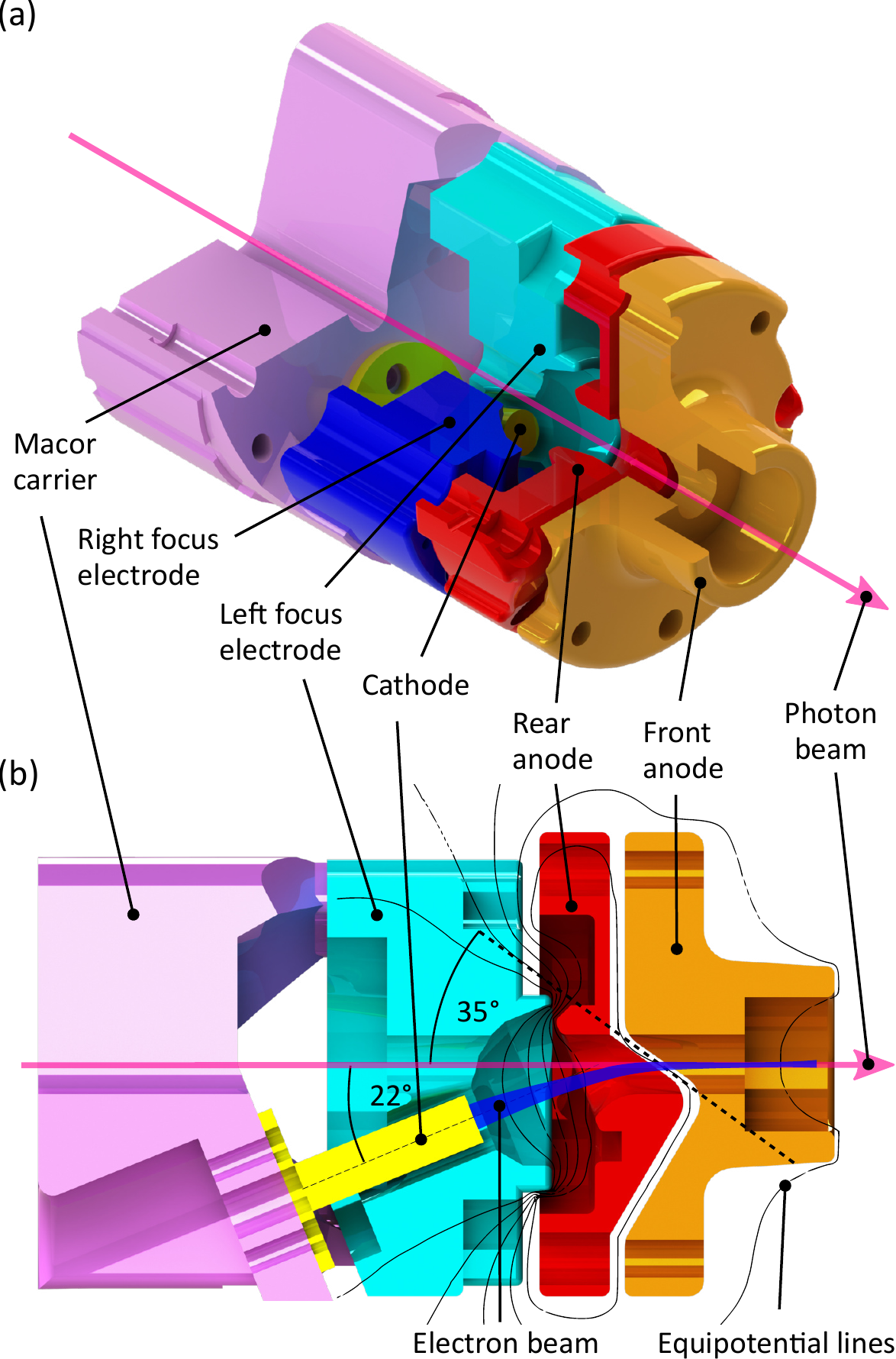}%
	\caption{Cross-sectional views of the off-axis electron gun\label{off-axis-gun}. The cathode (yellow) points at an angle of $22^{\circ}$ to the trap axis to allow an external photon beam to pass through the unobstructed central bore. The mirror-symmetric focus electrodes (light blue and blue) compensate for the drift of the electron beam due to the Lorentz force. Bending of the electron beam onto the trap axis is realized by splitting the anode into a rear (red) and a front (orange) electrode, cut by a $35^{\circ}$-plane with respect to the horizontal. Simulated electrostatic potential lines (black) and electron beam trajectories (blue) are also shown in (b). See text for further details.}%
	\label{oag1}
\end{figure}

\begin{figure*}
	\includegraphics[width=0.9\textwidth]{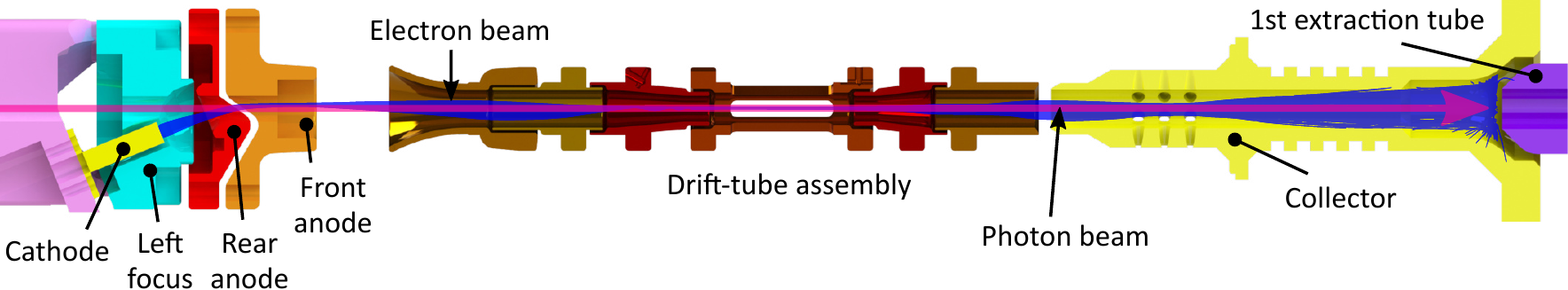}
	\caption{Cross section through the electrostatic elements of the PolarX-EBIT\label{off-axis-simulation}. Electron trajectories are shown in light blue. Due to the unobstructed trap axis, photon beams can pass through the EBIT, injected either from the gun or the collector side.}
	\label{OffAxisOverview}
\end{figure*}

To solve these problems we have developed a novel off-axis electron gun as an alternative to the on-axis gun and equipped PolarX-EBIT with it. Hence, a photon beam can propagate through PolarX-EBIT without any obstructions while being axially overlapped with the trapped HCIs.
For this reason, the cathode is separated from the trap axis by tilting it by an angle of $22\,^{\circ}$ with respect to the horizontal plane (see Fig.~\ref{off-axis-gun}(b)). Directly mounted on a precisely machined Macor insulator, the cathode is located 9.5\,mm in front of the anode. To optimize the deflection and focusing capability of the gun in the magnetic field, we simulated the electrostatic potentials and electron trajectories, using the software SIMION (see Figs.~\ref{off-axis-gun}(b) and \ref{off-axis-simulation} for visualization of such simulations).
For steering the electron beam onto the trap axis, the anode is split into two separate electrodes by a $35\,^{\circ}$-plane with respect to the horizontal plane at the intercept of the electron beam (see Fig.~\ref{off-axis-gun}(b)). Whereas the rear anode is used to define the extraction potential inside the gun, the front anode is used to bend the beam into the horizontal to direct it towards the trap center. Since the electrons emerge non-coaxially to the magnetic field lines, their trajectories are deflected sideways due to the Lorentz force. To compensate for this, the focus element surrounding the cathode is vertically cut into two mirror-symmetric electrodes (left and right focus) on which different potentials are applied. Additionally, the focus electrodes are also used, similarly as for the on-axis gun, to regulate the emission current. These electrodes are mounted on the insulating Macor carrier. Owing to the high temperature of the nearby cathode, the focus electrodes are made of molybdenum. The rear and front anodes, in turn, are mounted on the focus electrodes and the Macor carrier by alumina-ceramic rods as spacers. They are made of OFHC copper to distribute the possible heat load by scattered electrons.

The central bore of the off-axis gun along the trap axis is 4\,mm wide and, thus, in accordance with the requirements of typical photon beam diameters of less than 1\,mm. The alignment with the aim to maximize the overlap between the photon and the electron beam for a high signal rate is eased by, first, the short trap length of our compact EBIT and, second, the capability to image the photon beam after passing through the EBIT.

\subsection{Drift tubes}
\begin{figure}
\includegraphics[width=0.9\linewidth]{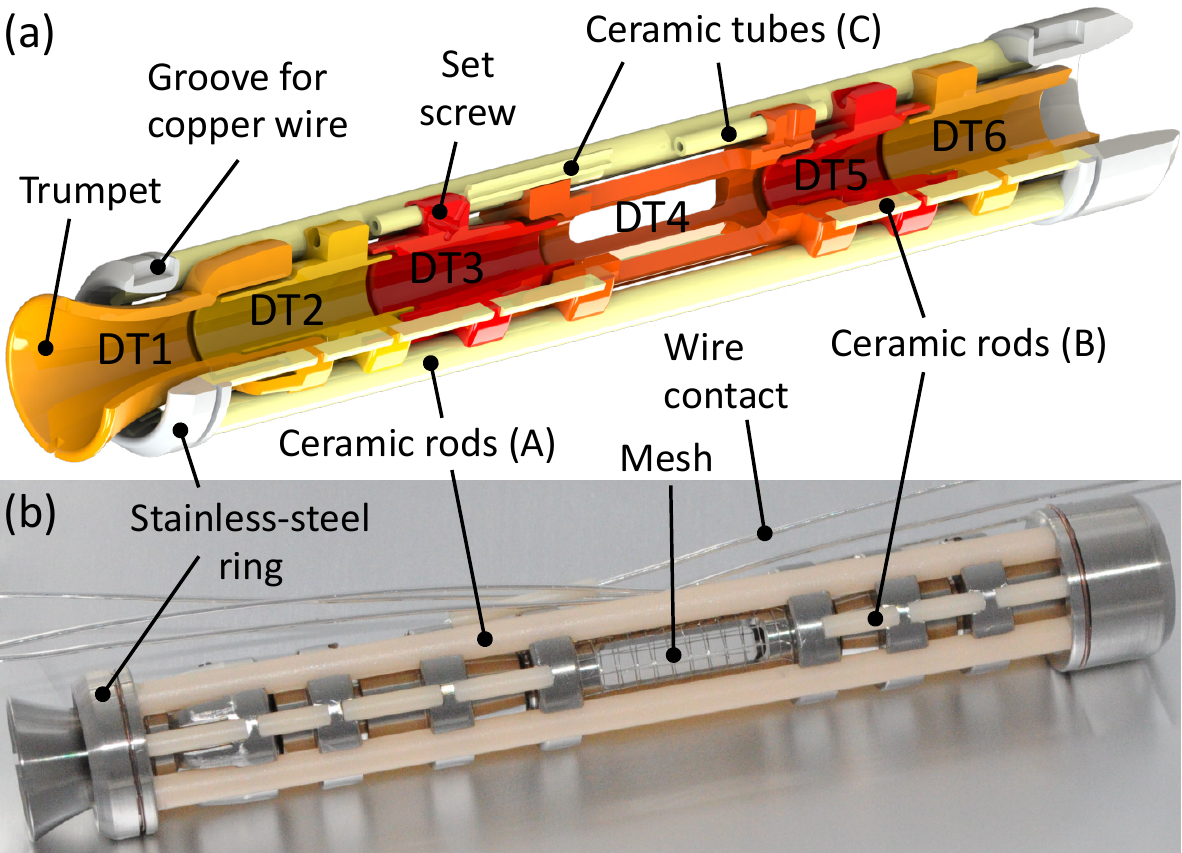}
\caption{Cross-sectional view (a) and photograph (b) of the drift-tube assembly\label{drift-tubes}. A set of six drift tubes is supported by alumina-ceramic rods and mounted between two stainless-steel rings. See text for further details.}%
\end{figure}

A set of cylindrical electrodes (drift tubes) accelerates and guides the electron beam towards the collector (see Fig.~\ref{El_sys}(b)) and also shapes the axial ion-trapping potential (see Fig.~\ref{El_sys}(c)). Six independent drift tubes (DT1 to DT6, manufactured of grade 5 titanium alloy) are stacked along four alumina-ceramic rods (A in Fig.~\ref{drift-tubes}, diameter of 3\,mm), which center these electrodes and terminate on either side in stainless-steel rings. Seven pairs of alumina-ceramic rods (B) (diameter of 2\,mm), precisely cut to the correct length, set the different distances between the drift tubes and the stainless-steel rings as well as electrically insulate them, accounting for appropriate creeping distances. Two twisted copper wires, guided within a groove in each of the rings and slots filed on the ceramic rods (A) on either side, keep this asymmetric 100 mm-long assembly together. After assembling, the trumpet of DT1 is screwed with its outside thread into the body of DT1. Flexible silver-plated copper wires are used for connecting the drift tubes by M1.6 set screws. The wires need to be preinstalled since their electrode contact is not accessible after installation. Thin ceramic tubes (C) are used for insulating them within the 16-mm bore of the central vacuum chamber. Next, the assembly is inserted into this bore and clamped together by two stainless-steel holders on either side, pushing against the outer stainless-steel rings of the assembly. The central trap electrode (DT4) occupies the position of maximum magnetic flux density and establishes a potential well of an effective trap length of 22\,mm together with DT3 and DT5 (see Fig.~\ref{El_sys}). Four slits of 2.5\,mm height by 16\,mm length are aligned with the four DN40CF cube-side ports. To reduce reach-through from the ground chamber potential, the electrode apertures are covered with an electro-formed stainless-steel mesh of 2\,mm$\,\times\,$2\,mm grid size. This geometry offers an opening angle of $58^\circ$ for each DN40CF port. One port is needed for HV feedthroughs and pumping. Another port is used for an atomic beam for trap loading. The two remaining ports are available for fluorescence detectors and spectrometers. Potentials applied to the first three drift tubes are tuned to optimize electron-beam focusing at the trap center (see Fig.~\ref{El_sys}(c)).

\subsection{Collector}
\begin{figure}
\includegraphics[width=0.9\linewidth]{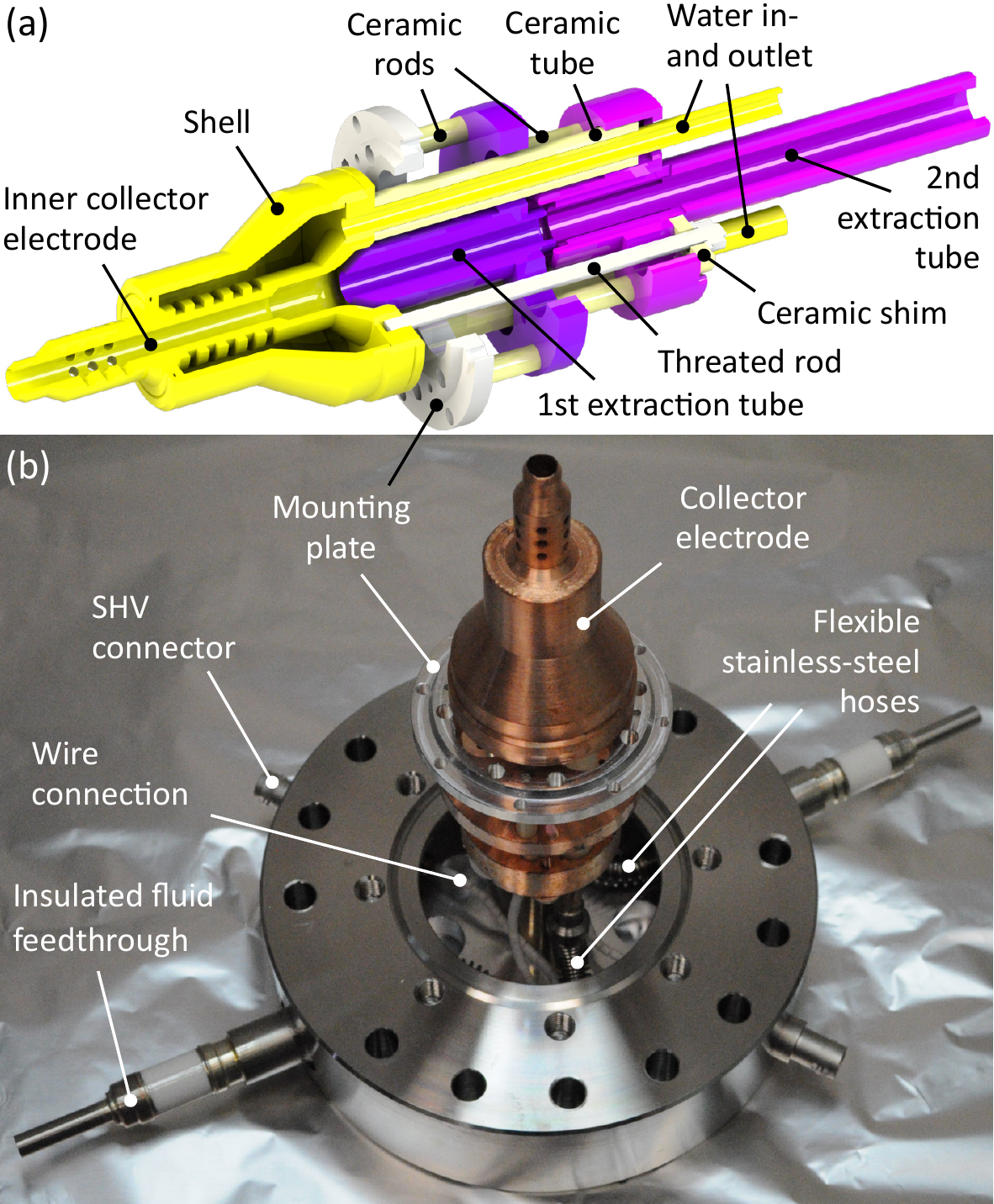}
\caption{Cross-sectional view (a) and photograph (b) of the collector\label{collector}. The water-cooled collector electrode (yellow in (a)), a mounting plate, and two extraction tubes are electrically insulated by ceramic elements against each other. Flexible stainless-steel hoses, brazed to hollow high-power feedthroughs, provide the collector with water. See text for further details.}%
\end{figure}

After passing the positively biased drift-tube section, the electron beam enters a region at ground potential and is therefore slowed down. Finally, it is dumped on the inside of a hollow water-cooled collector electrode (OFHC copper) (see Figs.~\ref{El_sys} and \ref{collector}). The magnetic field strength at the collector is much lower than at the trap center, and thus the electron beam expands again and hits the wall. Behind the collector electrode, two extraction electrodes are installed. The first one has to be biased to a more negative potential than the cathode to prevent the electron beam from passing through the collector. Furthermore, both tubes are also used as ion optics for HCI extraction.

The collector is made of two copper parts, the inner collector electrode and the outer shell, enclosing a volume for cooling water. These elements and the copper pipes for water inlet and outlet are electron-beam welded together. The collector, an aluminum mounting plate, and the two extraction tubes (OFHC copper) are stacked by using alumina-ceramic rods as spacers and two threaded rods to clamp the assembly together. Further ceramic shims and ceramic tubes around the water pipes are used for electrical insulation. Stainless-steel hydro-formed, braze-soldered flexible-bellow hoses connect the collector piping to insulated fluid feedthroughs. These feedthroughs lead the current deposited by the electron beam to the outside. The collector flange is mounted on its DN63CF side to the central vacuum chamber while it widens on the other side to DN100CF for attaching further extraction elements and a beamline. There, a 300\,l/s turbomolecular pump (TMP) can be installed to pump the collector section from the back. Inside the central vacuum chamber, the collector assembly is mounted with the mounting plate. Two safe-high-voltage (SHV) connectors are welded on the collector flange for biasing the extraction tubes. The electrical insulation of the collector also allows for biasing it, provided a protecting enclosure is installed. However, in the current setups it was operated very close to ground potential by connecting it through a 10\,$\Omega$ resistor or an ampere-meter measuring the current $I_{col}$ for evaluating the transmission $T = I_{col} / I_b$. This is a key indicator for the quality of the electron beam, with typical values of $T\approx 99\,\%$. The missing current lost to the electron-gun anode and the six drift tubes is monitored and minimized during voltage adjustment.

\subsection{Pumping and injection system}
A low residual gas pressure is an essential prerequisite for achieving and keeping high charge states. We use a cascaded TMP system to achieve an ultra-high vacuum (UHV) at the low $10^{-9}$\,mbar level: Four TMPs (preferably 70\,l/s for the electron gun, 70\,l/s for the trap, 300\,l/s for the collector and 70\,l/s for the injection system, or better) share downstream a common intermediate high-vacuum stage, which is pumped by a single 70\,l/s wide-range TMP, backed by an oil-free scroll pump. The use of this two-stage TMP system raises the compression ratio for H$_2$ by more than three orders of magnitude and prevents serious problems caused by a failure of either the intermediate TMP or one of the UHV TMPs. In the first case, the scroll pump can maintain the required backing pressure for the UHV TMPs. In the second one, the small intermediate TMP still manages to keep a vacuum in the 10$^{-6}$\,mbar range in the EBIT. Then, a vacuum-interlock safety system automatically turns off HV power supplies and the cathode heating unit. This level of protection is sufficient to prevent the cathode from permanent damage at the price of introducing the intermediate high-vacuum stage TMP. Additional protection is provided by a solenoid valve at the inlet of the scroll pump. It closes in case of a power failure and maintains, together with the forelines, a level of vacuum which is suitable for the continuing operation of the TMPs during their spin down. Again, the vacuum-interlock system immediately switches off the power supplies and also keeps them switched off when power is suddenly restored, up to their manual restart. In this way, the cathode has enough time to cool down, since this process takes many minutes. Hot-cathode ion gauges are installed in the various chamber parts to monitor the pressure.

For immediate operation of the HC-EBITs after a transport between laboratories or cities, the electron gun can be isolated with gate valves in UHV in its own chamber. This protects the cathode from degradation and avoids a time-consuming re-activation after arrival. Additionally, we have installed a non-evaporable getter (NEG) pump close to the gun for the PolarX-EBIT to maintain a UHV pressure for an extended period without electrical power.

The gas-injection system has two UHV pumping stages to deliver a very low-density atomic or molecular beam to the central drift tube in order to introduce the element of choice into the EBIT. An UHV needle valve is connected to the first UHV stage, pumped by the aforementioned 70\,l/s TMP. There, the pressure is typically in the range of 10$^{-6}$ to 10$^{-8}$\,mbar. An aperture of 2\,mm diameter allows the beam to enter the second UHV stage, which is pumped by the TMP attached to the central chamber. Moreover, Tip-EBIT uses an implementation of the wire-probe method\cite{elliott_wire_1995} in order to deliver minuscule quantities of heavy elements for ionization. The wire probe will be used for the study of rare isotopes.

\section{Operation and measurements}\label{meas}

\subsection{Commissioning}\label{commissioning}
All devices of the HC-EBIT type showed a similar behavior during commissioning. At the beginning, the brand-new cathode is carefully heated up in a stepwise manner to the required activation temperature over a couple of days, dubbed `conditioning'. This process causes a slow diffusion of the impregnants, driven by the high temperature and the concentration gradient, during which the barium aluminate reacts with some additives contained in the cathode, e.g., calcium oxide, and forms barium. Due to its low work function ($\approx 2$\,eV), this substance builds a very efficient emission layer at the exterior surface of the cathode. The cathode suppliers recommend not to exceed a certain heating current since the type `M' coating evaporates at too high temperature. At higher values the cathode itself can be damaged. However, due to the open gun structure, the coupling through thermal radiation to the environment is relatively good and we expect to be clearly below the temperature limit. Because of the initial heating of the cathode and its environment, the pressure, measured by the gauge which is installed at the gun chamber, approaches the 10$^{-7}$\,mbar range while the pressure measured at the collector remains in the 10$^{-9}$\,mbar range. Increased ion bombardment of the cathode emission layer due to residual gas pressure removes some of the emitting barium and counteracts the diffusion process of the barium from the tungsten-matrix reservoir -- initially limiting the maximum emission current.

After a few weeks of operation, the chemical-physical formation of the dispenser-cathode material has completed, the parts immediately surrounding the cathode have out-gassed, and the cathode heating can be reduced to prolong the lifetime, while maintaining the emission current. This process can be accelerated by baking the whole device while cooling the permanent magnets using the water-cooling of the protecting aluminum cartridges. After vacuum conditions have consistently improved (low 10$^{-9}$\,mbar pressure at the gun), currents of up to 80\,mA for the on-axis gun and up to 30\,mA for the off-axis gun could be achieved. Nonetheless, some of the commissioning measurements presented below were obtained at lower currents, either because of a higher residual gas pressure at the beginning or intentionally to achieve higher electron-beam energy resolution.

Electrical discharges between the gun electrodes can temporarily reduce or completely eliminate the emission layer. Reconditioning of the cathode layer after recovery of good vacuum conditions can be achieved, in most cases, in a matter of seconds or minutes. However, extended, strong discharges can completely sputter the emission layer or cover it with other materials from neighboring electrodes. Such an incident can require many hours or even days of reconditioning at higher cathode-heating current, until the contaminating layers have been evaporated and a fresh barium emission layer has formed through permeation out of the reservoir. More serious problems can arise if an accidental leak occurs. In such cases, an vacuum-interlock system should instantly turn off the cathode heating and the high voltages applied to all the electrodes. Depending on how high the pressure rises during such an event and how long a high cathode temperature was maintained before restoring good vacuum conditions, several days of slow emission-current recovery can be required. In the worst case, the damage to the cathode is permanent.

\subsubsection{On-axis electron gun}
\begin{figure}
\includegraphics[width=1\linewidth]{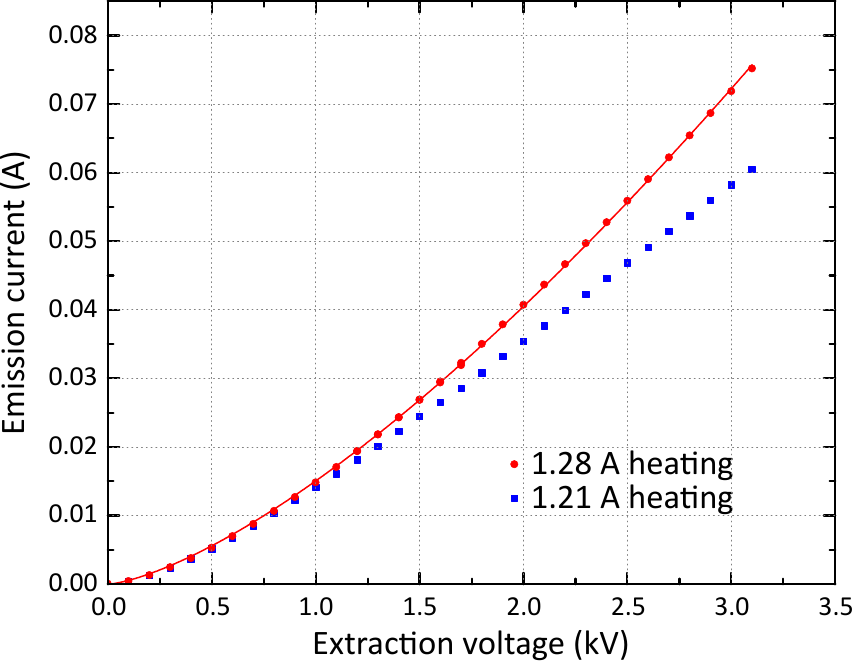}
\caption{Perveance measurement of the HC-EBIT on-axis electron gun\label{perveance}. For two different cathode-heating currents, the emission current is shown as a function of the extraction voltage, i.e., the voltage difference between anode and cathode while the focus electrode was kept on cathode voltage. A heating with 1.21\,A (blue squares) indicates emission not being fully space-charge limited. The fit function $I_b = a \, U^{c}$ yields the parameter $a= 0.78 (2) \times 10^{-6}$\, A/V$^c$ and $c = 1.428 (3)$ for the higher-temperature data (red circles).}%
\end{figure}
The Pierce-type electron guns operate in the space-charge limited regime. Appropriate high voltages have to be applied to the different electrodes. Typically, the focus electrode is initially biased to a more negative value than the cathode, thus completely blocking the emission. The anode voltage can be kept on ground or a low positive potential of only a few hundred volts. By increasing the focus voltage, emission slowly starts and the position of the electron gun can be adjusted aiming at maximum transmission all the way through the drift tubes to the electron collector by monitoring the emission current and the collector current. A rough positioning of the gun was always sufficient to immediately measure electron-beam current on the collector. In a second step, the current loss to the anode and the different drift tubes are minimized by fine positioning the gun and adjusting the drift-tube voltages. The reason for the anode current is essentially a partial reflection of the strongly focused electron beam by the strong magnetic-field gradient, hindering the electron beam from entering the drift-tube region. Consequently, the anode collects the largest part of the reflected beam, since it is biased more positively than any other part there. Therefore, the anode current is a proxy for the quality of both the geometric alignment and the voltage adjustments. Once these steps are completed, having a typical transmission of 99\,\% and a current loss to the anode and each drift tube well below 0.5\,mA (even at emission currents of $80$\,mA), the anode voltage can be increased to obtain the desired electron-beam current. Possibly, the gun position, the focus voltage and the drift tube voltages may need some readjustments then. The geometric alignment is very reproducible and hardly varies for a wide range of electron-beam energies and currents. However, for each beam energy adjustments of the drift-tube and electron-gun potentials may be needed. Generally, also these parameters are reliably reproducible and can be maintained at constant values during long measurements.

The perveance $P = I_b / U^{3/2}$ is an important quantity to describe the relation between beam current $I_b$ and extraction voltage $U$ for space-charge limited, charged particle beams. In particular, it is often used to qualify the performance of electron guns. According to Child-Langmuir's law\cite{child_discharge_1911,langmuir_effect_1923} it is essentially determined by two geometrical quantities for a given particle beam: the diameter of the extracted beam and the distance between cathode and anode. For our on-axis electron gun we expect a perveance on the order of 1\,$\mu$perv with the 3.4\, mm-diameter cathode. However, in practice this relation is not always fully applicable due to geometrical deviations and experimental conditions, for instance, in the presence of a magnetic field when the gun is operated in an EBIT. Fig.~\ref{perveance} shows the emission current as a function of the electron-gun extraction voltage, i.e., the difference between anode and cathode voltage while the focus electrode is operated at cathode potential. Fitting the data of $1.28$\,A cathode-heating current, the gun operating in the space-charge limited regime, we can confirm an electron-gun perveance on the order of 1\,$\mu$perv. Note that the electron-beam current of the HC-EBITs, adjusted and regulated by the anode and focus-electrode voltages, respectively, is widely independent of the electron-beam energy which is determined by the voltage difference between trap electrode and cathode. The measurement was carried out with 3.6\,keV beam energy, for instance.

\subsubsection{Off-axis electron gun}
In first tests of the off-axis gun, the electrostatic deflection of the emitted electron beam was analyzed in a non-magnetic environment. 
In this setup, the electrons hit a phosphor-coated screen mounted in front of the gun. 
By changing the potentials of the electrodes, the electron beam could be steered along the horizontal and vertical direction independently. 
The focussing of the beam could be adjusted by changing the bias voltage of the focus electrodes. 
Then, the off-axis gun was mounted on an XYZ-manipulator and installed in the PolarX-EBIT. 
The manipulator allows positioning of the gun at the magnetic-field minimum, a crucial requirement.
Starting with electrode potentials and a gun position optimized by simulations, the cathode emitted already electrons and the beam could be partially transmitted through the EBIT to the collector. 
After a manual optimization, stable beams with electron losses below 1\,\% were achieved.

The Lorentz deflection strongly depends on the gun position. 
If the cathode is misaligned, compensating the beam drift calls for a voltage difference between the two focus electrodes of up to 50\,V. 
Its optimal value does not strongly depend on the voltage difference between cathode and anode. 
The potential between front and rear anode, required for proper electron-beam steering, is in good agreement with the simulations. 
Positive voltages at the front anode are approximately half of those applied to the rear anode, a ratio which is not much affected by the electron-beam current.
However, if the voltage difference between cathode and rear anode is increased, i.e., essentially the beam energy is increased inside the gun, the voltage applied to the front anode also needs to be increased.

PolarX-EBIT was operated with stable electron beams from a few hundred eV up to 8000\,eV energy at the trap center. The cathode could be biased down to -4000\,V, the limit of the cathode power supply. Stable space-charge limited currents of up to 30\,mA were achieved, limited by the maximum voltage difference between cathode and rear anode while a proper deflection of the electron beam onto the trap axis was still possible.

\subsection{HCI extraction}\label{IE}

\begin{figure}
\includegraphics[width=1\linewidth]{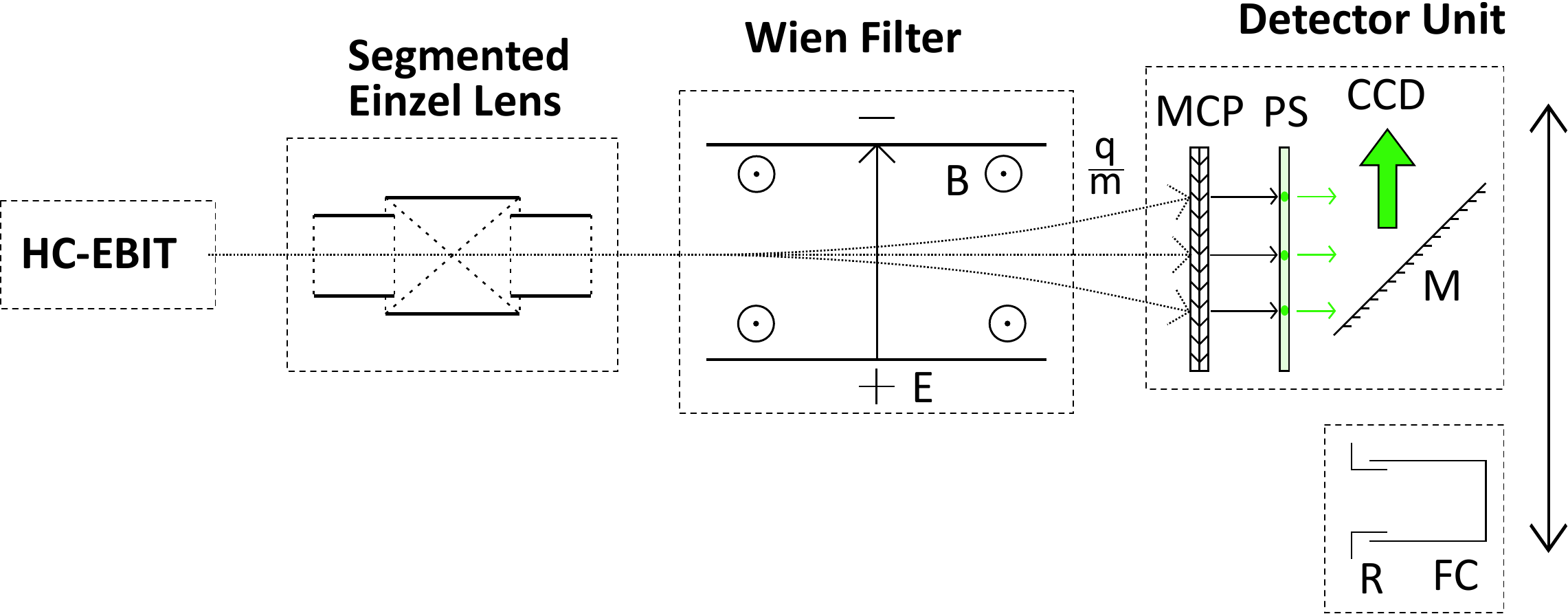}
\caption{HCI beamline setup\label{beamline}. B: Magnetic field, CCD: Charge-coupled device camera, E: Electric field, FC: Faraday cup, M: Mirror, MCP: Microchannel plate, PS: Phosphor screen, R: Repeller electrode.}%
\end{figure}

\begin{figure}
\includegraphics[width=1\linewidth]{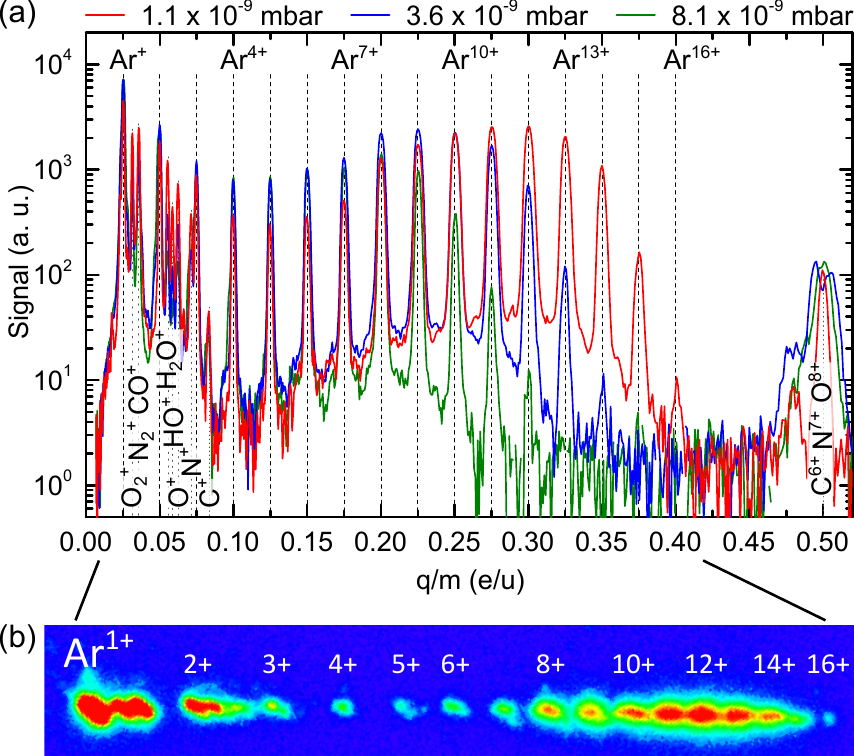}
\caption{(a) Microchannel-plate detector signal after charge-state separation (by Wien filter) of extracted Ar HCI bunches using a 4\,mA, 2.8\,keV electron beam at different EBIT chamber pressures\label{ArExtPul} set by neutral Ar injection. Top labels: Ar charge states. Bottom labels: molecular and atomic ions from residual gas. (b) Charge-state distribution of a single Ar ion bunch impinging on the microchannel-plate detector. All charge states up to $q = 16$ are present at once, plus molecular and atomic ions from residual gas.}%
\end{figure}

\begin{figure}
\includegraphics[width=1\linewidth]{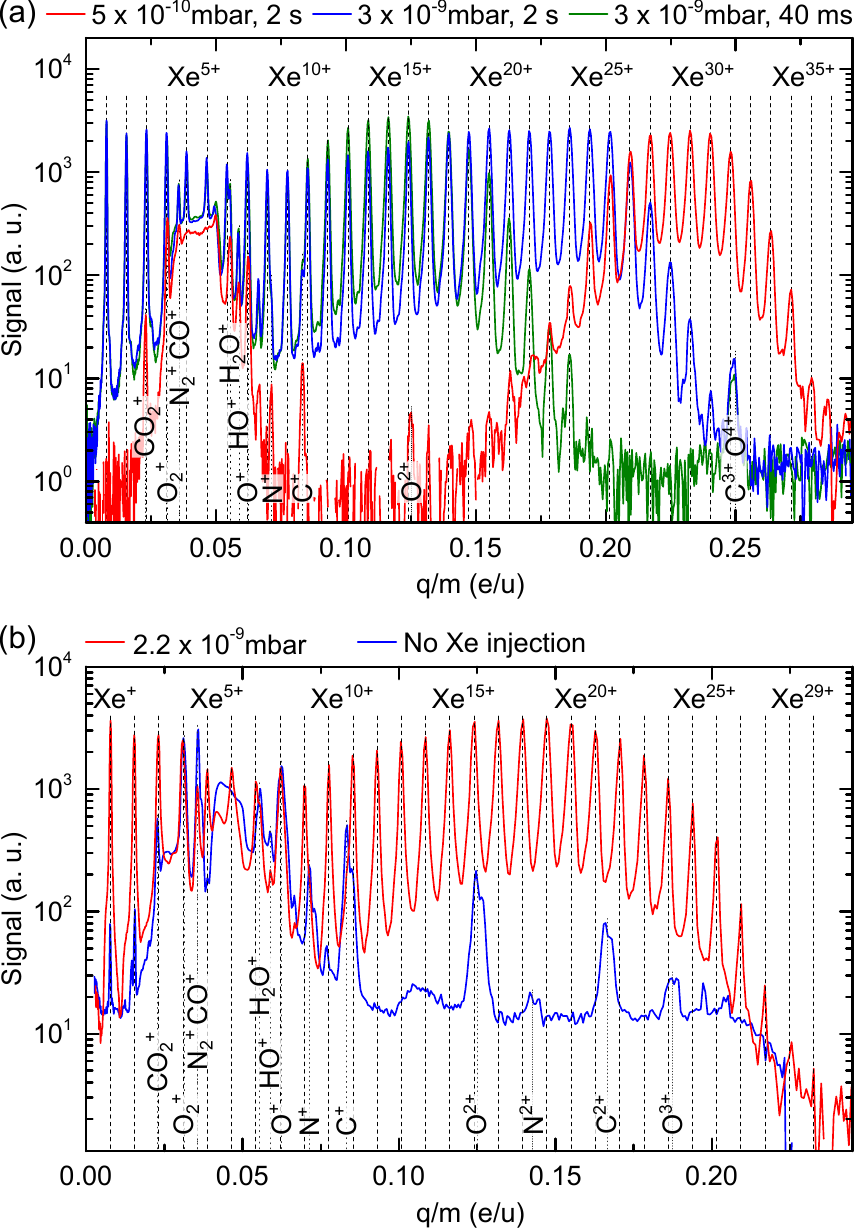}
\caption{Charge-state distribution of extracted $^{129}$Xe\label{Xe}. (a) Pulsed extraction for different EBIT pressures and breeding times with a 4.6\,mA, 3\,keV electron beam. (b) Continuous extraction with and without Xe injection employing a 4\,mA, 2\,keV electron beam. Top labels: Xe charge states. Bottom labels: molecular and atomic ions from residual gas.}%
\end{figure}

To test ion breeding and extraction, we installed the prototype of the HC-EBITs at a beamline with charge-state separation and detection. This device has a slightly lower magnetic field of 0.74\,T at the trap center, uses an older gun design, and delivered electron-beam currents of a few mA at the time of the tests. Fig.~\ref{beamline} illustrates the beamline setup. The HCIs leave the EBIT through an extraction electrode behind the collector (cf. Fig.~\ref{El_sys}(b)) and pass through a special, segmented einzel lens, dubbed Sikler lens, similar to our previous design\cite{simon_photoionization_2009,ginzel_deceleration_2010}. The Sikler lens allows focusing and steering of the ion beam in spite of its short length. It forms a collimated beam towards a Wien-type velocity filter for separation according to the charge-to-mass ratio $q/m$. The electric field of the Wien filter is scanned for $q/m$ analysis using alternatively either a Faraday cup measuring ion currents or a position-sensitive detector comprised of a microchannel plate (MCP) and a phosphor screen (PS), imaged onto a charge-coupled device camera (CCD).

For operating the EBIT in pulsed mode, an axial trap depth of up to 100\,V is applied. At the cost of a high ion temperature, this allows to breed high charge states. By rapidly switching the trapping potential, an HCI bunch is released. We tested this mode with argon (see Figs.~\ref{ArExtPul}(a) and \ref{ArExtPul}(b)) and isotopically enriched \mbox{xenon-129} (see Fig.~\ref{Xe}(a)). We also carried out continuous extraction (\lq leaky\rq{} mode) with the prototype using a shallow axial trapping potential. Hereby, ions of the hot tail of the Maxwell-Boltzmann distribution overcome the axial trapping-potential barrier continuously. Fig.~\ref{Xe}(b) shows results for $^{129}$Xe.
Table~\ref{Ext_tab} summarizes the main results of these experiments. The HCI charge-state distribution for the injected elements, and further peaks from molecular and atomic ions arising from residual gas, are visible in the figures. A superimposed, diffuse background between 0.03 and 0.06 e/u corresponds to ions not originating from the trap-center potential. With a 4\,mA, 2.8\,keV electron beam, Fig.~\ref{ArExtPul}(a) shows all Ar charge states up to the He-like Ar$^{16+}$, the highest accessible charge state for Ar at that electron-beam energy. Fig.~\ref{ArExtPul}(b) shows a CCD image of a single Ar HCI bunch spatially separated at low Wien-filter dispersion displaying simultaneously all produced charge states. This allowed to quickly assess the whole charge-state distribution at once. Extraction of much heavier $^{129}$Xe HCIs yielded charge states of up to $q=37$ (pulsed at 4.6\,mA, 3\,keV electron beam, see Fig.~\ref{Xe}(a)) and $q=29$ (leaky mode operation at 4\,mA, 2\,keV electron beam, see Fig.~\ref{Xe}(b)), respectively, at sufficiently low pressures. Faraday-cup measurements yielded 100\,pA of the total extracted ion current for the charge-state distribution shown in red in Fig.~\ref{Xe}(b).

\subsection{Dielectronic recombination}\label{DR}

\begin{figure}
\includegraphics[width=0.8\linewidth]{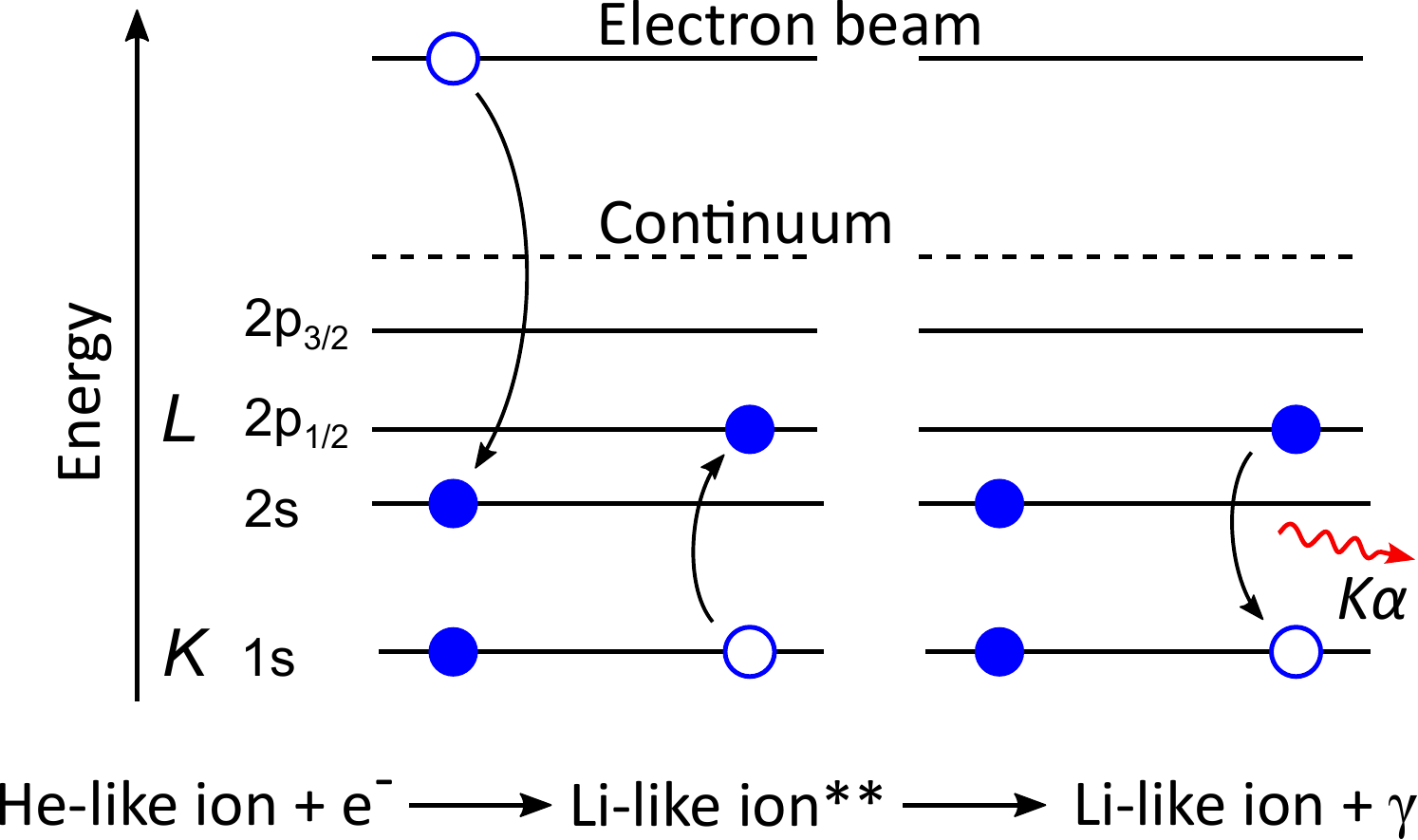}
\caption{Simplified energy scheme for resonant $KLL$ dielectronic recombination of a He-like ion\label{DR_sketch}. A free electron from the beam is captured into the $L$ shell while a second, bound $K$-shell electron is also excited into the $L$-shell. The intermediate hole state of the resulting Li-like ion decays through $K\alpha$ emission.}%
\end{figure}

\begin{figure}
\includegraphics[width=1\linewidth]{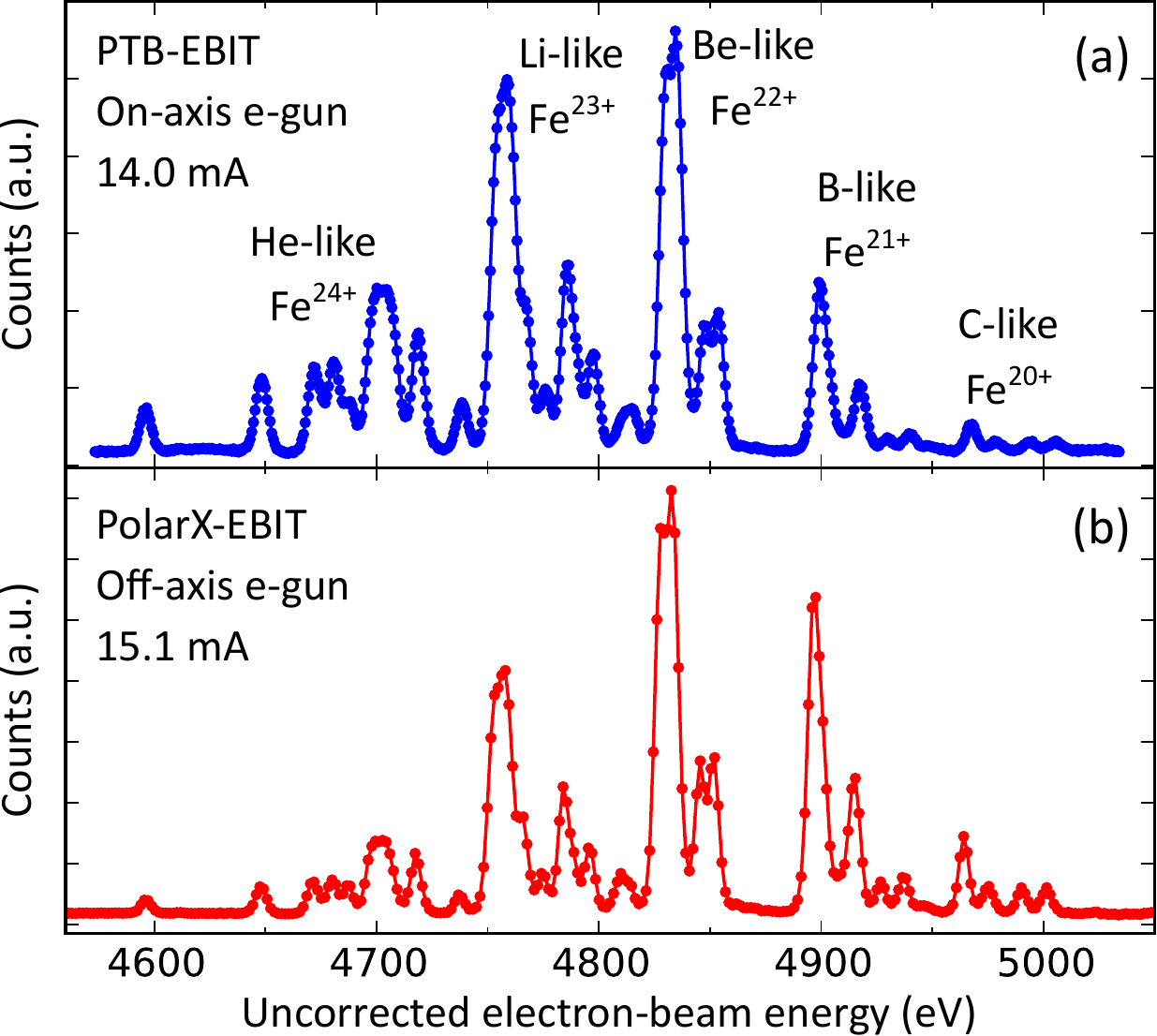}
\caption{Dielectronic recombination of iron ions\label{Fe}. The $K\alpha$ photons produced by the de-excitation of the intermediate hole states, populated by resonant dielectronic capture, were counted with a high-purity Ge detector while scanning the electron-beam energy. Shown are raw data, uncorrected for the space-charge potential and the influence of adjacent potentials. Since the $KLL$ dielectronic recombination of Fe is well known\cite{beilmann_major_2013}, the charge states can be unambiguously assigned. The upper panel shows a measurement with PTB-EBIT and the on-axis electron gun with a 14\,mA electron beam, while the lower panel shows a measurement of PolarX-EBIT equipped with the off-axis electron gun operated with 15.1\,mA.}%
\end{figure}

Dielectronic recombination\cite{massey_properties_1942} (DR) is a resonant interaction process of a free electron with an ion. The former one, having a kinetic energy of $E_{kin}$, is captured into an open shell of the latter one with binding energy $E_B$ while a second, bound electron is excited to a higher level with energy difference $\Delta E$ by the released energy $E_{kin} + E_B$ -- fulfilling the resonance condition $\Delta E = E_{kin} + E_B$. A resulting excited, short-lived intermediate state is stabilized either by autoionization or radiative decay. $KLL$ DR (notation akin to the Auger nomenclature referring to the  electronic shells involved) is illustrated in Fig.~\ref{DR_sketch} for an initially He-like ion. Observation of $KLL$ DR is an excellent diagnostic of the charge-state distribution\cite{takacs_diagnostic_2015}, since the energies of charge-state-resolved resonances are often well known. Hence, we carried out DR measurements on iron with the HC-EBITs, injecting an Fe-containing organometallic compound. While scanning the electron-beam energy over the $KLL$ resonances, a high-purity germanium detector counted the $K\alpha$ photons, which were emitted by stabilizing the intermediate state. Fig.~\ref{Fe} presents results for two different HC-EBITs, the PTB-EBIT featuring the on-axis electron gun (upper panel) and the PolarX-EBIT utilizing the novel off-axis electron gun (lower panel) with electron-beam currents of 14\,mA and 15.1\,mA, respectively. The appearing resonances give clear evidence of high charge states up to He-like Fe$^{24+}$. The HC-EBITs reach a remarkably high electron-beam energy resolution of $E/ \Delta E > 1500$ for low electron-beam currents, even improving on that of previous work using evaporative cooling in cryogenic EBITs\cite{beilmann_prominent_2011,beilmann_major_2013}. Arguably, this could be interpreted as a consequence of the higher residual gas pressure in the room-temperature HC-EBITs which provides low-atomic number HCIs for evaporative cooling by default.

\begin{table}
\caption{Settings and results of HCI extraction obtained with the prototype\label{Ext_tab}}
\begin{tabular}{llcccc}
\hline
\hline
Current &Energy &Mode&Element&Most abundant & Highest \\
\text{(mA)}& \text{(keV)}& & &charge state &charge state \\
\hline
4 & 2.8 & pulsed & Ar & 12 & 16\\
4.6 & 3 & pulsed & Xe & 30 & 37\\
4 & 2 & leaky & Xe & 19 & 29\\
\hline
\hline
\end{tabular}
\end{table}

\subsection{Electron beam properties at the trap}\label{EB}
\subsubsection{On-axis electron gun}
The size of an electron beam, compressed in a coaxial magnetic field $B$, can be described by the Herrmann radius $r_H$\cite{herrmann_optical_1958}. It contains 80\,\% of the electron-beam current $I_b$, emitted by a cathode of radius $r_c$ at a temperature $T_c$ and in a residual magnetic field of $B_c$.
 \begin{equation}
 \label{eq:r_h}
 	r_H = r_B \sqrt{
 		\frac{1}{2} + \sqrt{
 			\frac{1}{4} +
 			\frac{8 m_e k_B T_c r_c^2}
	 			{ e^2 B^2 r_B^4} + 
	 		\frac{ B_c^2 r_c^4 }
		 		{ B^2 r_B^4}
 			}
 		},
 \end{equation}
where $r_B$ is the Brillouin radius
 \begin{equation}
 r_B = \sqrt{ \frac{ 2 m_e I_b } {\pi \epsilon_0 v_z e B^2} }
 \end{equation}
with the electron mass $m_e$, charge $e$, and velocity $v_z$, as well as the Boltzmann constant $k_B$.

Accordingly, in the magnetic field of 0.86\,T at the trap center, as shown in  Fig.~\ref{El_sys}(a), the electron-beam radius is compressed down to 72\,$\mu$m, assuming a cathode temperature of 1400\,K, a cathode diameter of 3.4\,mm, an electron-beam current of 80\,mA, an electron-beam energy of 5\,keV, and a residual magnetic field at the position of the cathode of about or even below 100\,$\mu$T. The latter one is a reasonable assumption, since the structure of the magnetic field causes a true minimum where the field direction reverses. The minimum is only slightly shifted by the earth's magnetic field or a weak magnetic background field in the laboratory. Photographic images of the ion cloud under electron-beam excitation (as displayed on the inset of Fig.~\ref{Design}) are, in spite of their limited spatial resolution, consistent with a radius below 150\,$\mu$m. Based on the Herrmann radius of 72\,$\mu$m, and also guided by the ionization efficiency, we estimate an approximate current density of 500\,A/cm$^2$ and an electron density of $7\times 10^{11}$\,/cm$^3$ at 5\,keV beam energy.

The electron-beam energy was limited by the power supplies to 10\,kV so far. Higher values of up to 15\,kV are likely to be possible in the future with other supplies and possibly exchanging HV feedthroughs. If required, modifications of the designs and lengths of the insulating parts can be implemented to increase creeping distances.

The energy shifts as a function of the electron-beam current observed in DR measurements by the space-charge potential of the electron beam and trapped ions at a trap depth of 10\,eV were measured to be 0.397(8)\,eV/mA at about 2.2\,keV electron-beam energy of argon $KLL$ DR and 0.505(3)\,eV/mA at about 4.8\,keV of iron $KLL$ DR.

\subsubsection{Off-axis electron gun}

The observation of DR in argon at an electron-beam current of 10\,mA while varying the nominal trap depth, i.e., the difference of the applied voltages of the central trap electrode and the two adjacent drift tubes, allows for a coarse estimate of the electron-beam radius through the electronic space-charge potential $\Phi_e$, though $\Phi_e$ is partially compensated by the positive space-charge potential $\Phi_i$ of the trapped ions. The sum of these two contributions is the total space-charge potential $\Phi_{tot} = \Phi_e + \Phi_i$ and reduces the nominal electron-beam energy, defined by the acceleration voltage applied to the electrodes.
Solving Poisson's equation by assuming, for simplicity, an infinitely long electron beam propagating coaxially to the trap axis with a top-hat charge density, the electronic space-charge potential only depends on the separation $r$ from the axis (see also \cite{penetrante_evolution_1991,currell_physics_2005}) and is given by
\begin{equation}
\label{eq:sc1}
\Phi_e(r) =
\Phi_0\left(\frac{r^2}{r_e^2}+\ln\frac{r_e^2}{r_D^2}-1\right)
\end{equation}
for $r\leq r_e$ and
\begin{equation}
\label{eq:sc2}
\Phi_e(r) = \Phi_0\ln\frac{r^2}{r_D^2}
\end{equation}
for $r\geq r_e$, where $r_e$ is the electron-beam radius, $r_D$ is the inner radius of the cylindrical drift tube surrounding the beam, and
\begin{equation}
\Phi_{0} = \frac{I_b}{4 \pi \epsilon_0 v_z}
\label{eq:phi_0}
\end{equation}
is the potential difference between the edge of the electron beam and the trap axis. $I_b$ is the electron-beam current, $v_z$ the electron's velocity, and $\epsilon_{0}$ the vacuum permittivity.

The total space charge $\Phi_{tot}$ was determined by measuring the shifts in the nominal electron-beam energy, at which the DR resonances appear, as a function of the electron-beam current. A value of 0.462(9)\,eV/mA was obtained\cite{kuhn_inbetriebnahme_2017}, yielding a negative value of $\Phi_{tot} = - 4.6$\,eV for a 10\,mA electron beam.

The space-charge contribution $\Phi_i$ of the trapped ions, at this constant electron-beam current of 10\,mA, could be estimated by increasing the nominal trap depth from a trapping condition to its inversion and even higher, to, first, gradually reduce the number of trapped ions and, then, push transient ions out of the trap volume. In addition to the trap depth, the electron-beam energy was repeatedly scanned as the second parameter of this measurement to cover the argon $KLL$ DR resonances. Fig.~\ref{TrapDepthScan} shows the $K\alpha$ fluorescence as a function of both, the nominal trap depth and the nominal electron-beam energy, calibrated with atomic structure calculations using the Flexible Atomic Code\cite{gu_flexible_2008}.
The loss of positive charges when flatten the trap, resulting in a change of the space-charge potential and, therefore, in a shift of the resonance positions on the electron-beam energy axis, is not very large until a nominal trap depth of $+13(2)$\,V when the trapped HCIs begin to escape along the trap axis. From here, the space-charge compensation is noticeably decreased and a significantly higher nominal electron-beam energy is needed to fulfill the DR resonance condition.
At a nominal trap depth of $+250$\,V, only a few ions are still present and the $K\alpha$ fluorescence is very weak.
The difference of the required nominal electron-beam energy between the beginning of emptying the trap at the nominal depth of $+13$\,eV and the end of this process is determined to a value of 22(2)\,eV. This is our estimate for the space-charge contribution $\Phi_i$ of the ions at 10\,mA electron-beam current. By subtracting this value from $\Phi_{tot} = - 4.6$\,eV and taking an electrostatic reach-through of 1.2\,V into account, we can infer the electronic space-charge potential $\Phi_e \approx -25.4$\,eV. However, the total space-charge potential within the electron-beam radius $r_e$ is essentially flat due to the presence of the compensating ions, and therefore $-25.4$\,eV is our assumption for $\Phi_e(r_e)$. Using Eq.~\ref{eq:sc2}, we calculate an upper bound for the electron-beam radius of $r_b \leq 74(30)\,\mu$m. Using Eq.~\ref{eq:r_h} from Herrmann's theory, we can confirm operating the cathode in a residual magnetic field on the order of $B_C \approx 1\,mT $. This is in good agreement with the simulated and measured magnetic field, considering the off-axis geometry and a possible misalignment.
According to Herrmann's theory, the radius of such a compressed electron beam only increases by less than 2\,$\mu$m when increasing the electron-beam current to 30\,mA. Consequently, we expect achievable current densities of more than 170\,A/cm$^2$ with a 30\,mA electron beam of the off-axis electron gun.

\begin{figure}
\includegraphics[width=1\linewidth]{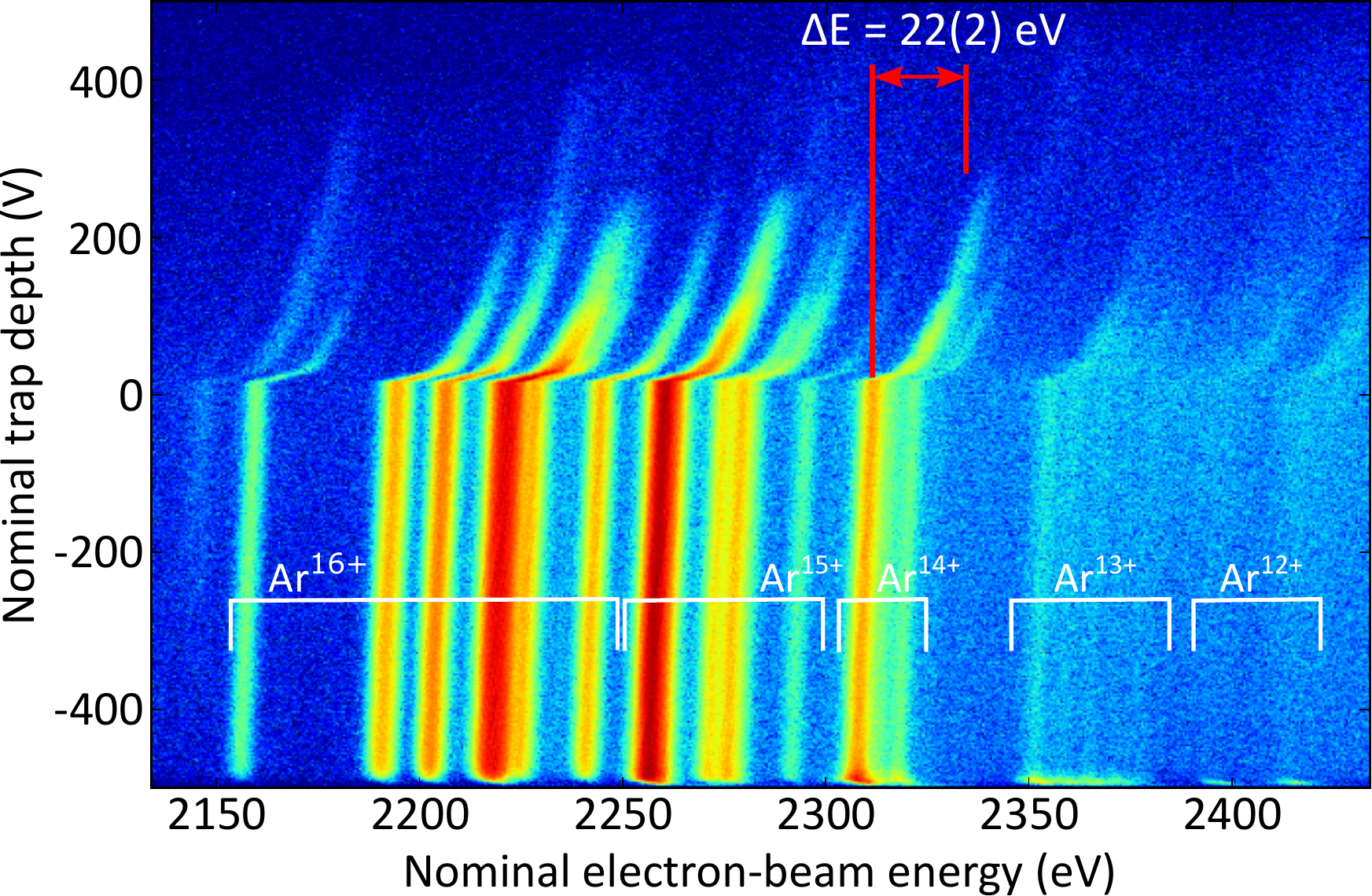}%
\caption{Dielectronic recombination of argon ions\label{TrapDepthScan} with constant 10\,mA electron-beam current as a function of the nominal electron-beam energy, defined by the applied acceleration potentials, and the nominal trap depth for storing the ions, i.e., the difference between the applied voltages of central trap electrode and adjacent drift tubes. The resonances are assigned to their corresponding charge states. Depending on the trap depth, the number of stored ions and, consequently, their positive space-charge contribution change. Hence, the nominal electron-beam energies at which the resonances appear shift. At $+13$\,V nominal trap depth, the trap starts losing ions, strongly affecting charge-state distribution, resonance intensities and space-charge compensation. The space-charge contribution of the ions was determined to 22(2)\,eV. See text for further details.}%
\label{TrapDepthScan}
\end{figure}

\subsection{Resonant photoexcitation}\label{bessy}
\begin{figure}
\includegraphics[width=1\linewidth]{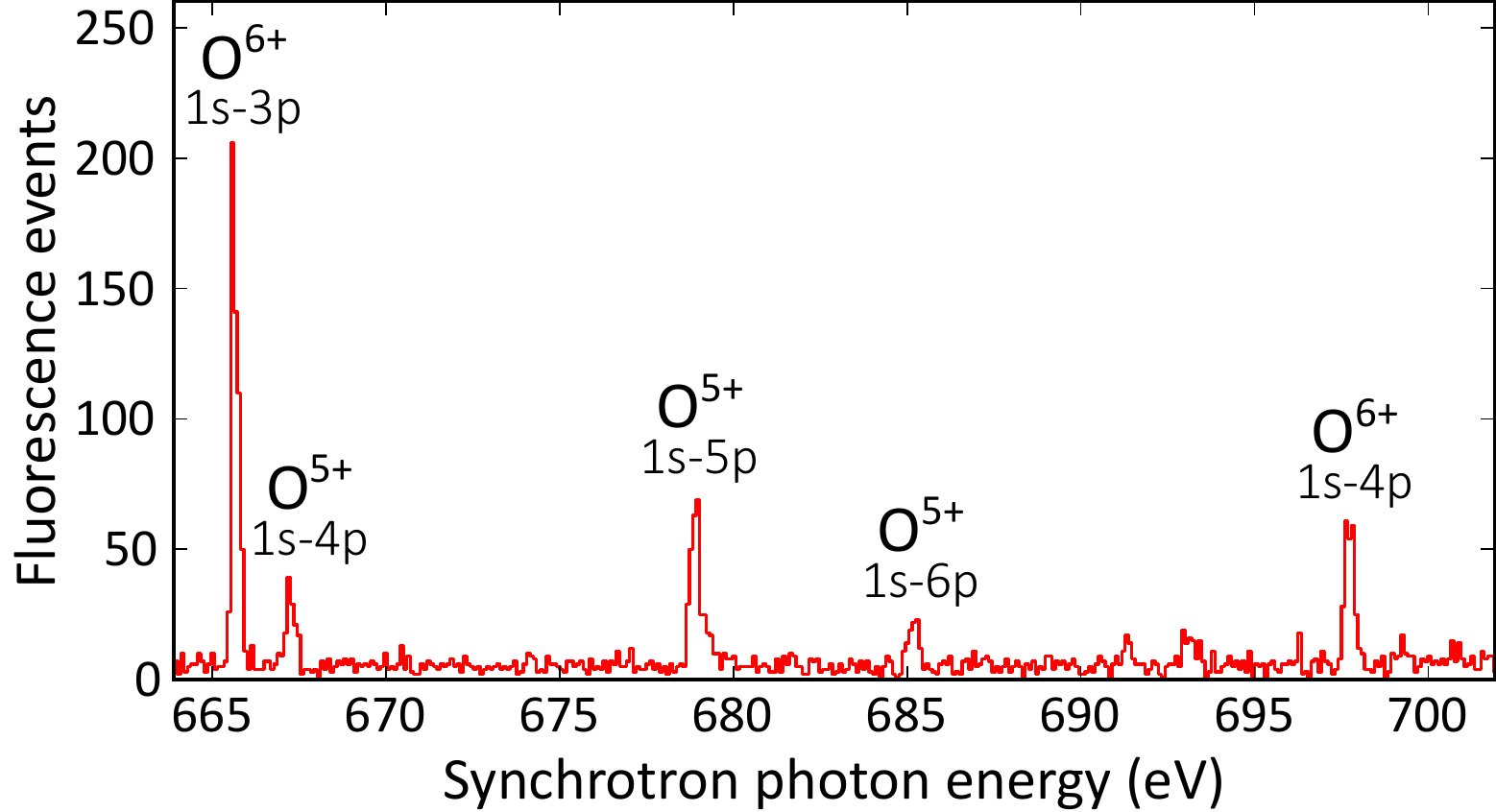}
\caption{Resonant photoexcitation of electronic transitions in highly charged oxygen\label{PE_spectrum}. The number of fluorescence events, recorded with two silicon drift detectors, is shown as a function of the photon beam energy. Five resonances of He- and Li-like oxygen are observed.}%
\end{figure}

Operating the transportable FLASH-EBIT\cite{epp_soft_2007} at FELs or synchrotron light sources has demonstrated the feasibility of resonant photoexcitation of electronic transitions in trapped HCIs by employing the laser spectroscopy technique in the XUV and x-ray regime\cite{epp_soft_2007, simon_photoionization_2010, epp_x-ray_2010, simon_resonant_2010, bernitt_unexpectedly_2012, rudolph_x-ray_2013, epp_single-photon_2015, steinbrugge_absolute_2015}. Tunable high-resolution monochromators allow high-precision measurements of transition energies and natural line widths. Moreover, such light sources provide femtosecond pulses. The PolarX-EBIT, with its novel off-axis electron gun, has been designed for the purpose of those experiments. We operated it at the BESSY II synchrotron in Berlin, providing a cloud of highly charged oxygen ions as target for the monochromatized x-ray beam of beamline U49/2-PGM1. PolarX-EBIT was set up with the electron beam collinear to the photon beam and was equipped with silicon drift detectors to detect the x-ray fluorescence signal of the HCI cloud. After coarse adjustment of the EBIT position using a YAG crystal and a phosphor screen, the spatial overlap of ion cloud and photon beam was optimized maximizing the fluorescence yield from the resonantly excited line $w$ (following the nomenclature of Gabriel\cite{gabriel_dielectronic_1972}) of He-like oxygen. A typical diameter of the ion cloud of a few hundred micrometers and an even smaller focus size of the photon beam necessitated a positioning accuracy on the order of 100 micrometers. By resonantly exciting electronic transitions in He-like O$^{6+}$ and Li-like O$^{5+}$, as shown in Fig.~\ref{PE_spectrum}, we successfully demonstrated the application of a room-temperature EBIT for high-resolution x-ray laser spectroscopy. Details on the measurement campaign and its results will be published elsewhere.

\section{Applications}\label{Appl}

Compared to conventional designs based on a superconducting magnet, requiring either a supply of liquid He or the continuous operation of cryocoolers, an HC-EBIT has a much reduced cost of construction and operation. It achieves, in spite of its compact size, the functionality of a traditional superconducting EBIT in many aspects. A large solid angle for detectors compensates for the lower electron currents and shorter trap dimensions to some degree. Mechanical vibrations from cryocoolers are eliminated. For many experiments in which an electron-beam energy below 10\,keV and current below 100\,mA are sufficient, an HC-EBIT can be employed more conveniently than its larger relatives. At the presently tested parameters one can expect to produce and study up to H-like iron, and He-like or Li-like ions up to Xe.

Once adjusted for an experiment, the HC-EBITs have demonstrated stable operation over months without readjustment. This renders these EBITs particularly attractive for the use as a pulsed or continuous HCI source. For instance, the prototype already successfully delivers HCIs for the Penning-trap experiment ALPHATRAP\cite{sturm_high-precision_2017} aiming at $g$-factor determination of the bound electron, PTB-EBIT will provide HCIs for a cryogenic Paul trap to carry out quantum logic spectroscopy testing HCI-based optical atomic clock applications, and Tip-EBIT is equipped with a wire-probe\cite{elliott_wire_1995} for rare-isotope injection to supply those HCIs to the Penning trap PENTATRAP\cite{repp_pentatrap:_2012} where precision mass measurements will be performed.

For spectroscopic applications, the HC-EBITs offers excellent optical access with a large opening angle along the trap axis. Optical, VUV, and x-ray detectors as well as optics for spectrometers, all can be mounted at close distance of only a few centimeters from the HCI cloud. This proximity and the subsequently possible large solid angles are more troublesome to achieve in superconducting EBITs due to the thermal shielding of the cryogenic parts. Similarly, it seems easier to adopt other EBIT techniques like the wire-probe method\cite{elliott_wire_1995} for rare-isotope studies. If the desired charge state can be achieved and signal intensity is sufficing, the seemingly disadvantageous low electron-beam current provides a higher resolution since the electron-beam energy spread $\Phi_0$ due to the electronic space charge (see Eq.~\ref{eq:phi_0}) is reduced and the HCIs are less heated, resulting in a reduced Doppler broadening. 

Moreover, PolarX-EBIT, now installed at the PETRA III synchrotron, was specifically developed for measurement campaigns at synchrotron or FEL facilities, providing HCIs as target for x-ray photons. Dedicated to this purpose we have designed and built a unique off-axis electron gun which allows for clear through access along the axially extended HCI cloud. An external photon beam can pass through the EBIT without being blocked and being available for a downstream experiment, facilitating operation of the EBIT in a parasitic mode with negligible transmission loss. Photon-energy calibration based on HCIs and photon-polarization diagnostics through the anisotropic angular emission become possible with the potential of providing an atomic absolute wavelength standard in this spectral region. Transport and installation of these compact machines are much more convenient than that of a large superconducting EBIT such as FLASH-EBIT \cite{epp_soft_2007}, and space requirements at a photon beamline are significantly relaxed.

\section{Conclusion}\label{Con}

In this paper we have introduced the Heidelberg Compact EBITs (HC-EBIT) as a novel class of compact room-temperature devices. With possible electron-beam energies of up to 10\,keV, more than 80\,mA electron-beam current, radial and axial access to the trapped HCIs, the design has proven its suitability for studies of H-like HCIs up to Fe, He-like and Li-like HCIs up to Xe, as well as all intermediate charge states of heavy elements. Featuring the novel off-axis gun, the operation at synchrotron and FEL light sources allows for innovative experiments with advanced photon-beam diagnostics based on atomic systems.

The HC-EBITs should provide reliable access to HCIs for a wide range of experiments, requiring much less expertise for its use, greatly reducing the investment and cost of operation. Three devices, PTB-EBIT, PolarX-EBIT, and Tip-EBIT, as well as the prototype are in operation. Three more are currently under construction. The HC-EBITs have already shown a performance competitive with fully-fledged superconducting EBITs in various applications. We are keen to share our design with other research groups to facilitate new applications of HCIs in, e.~g., atomic physics, astrophysics, surface science, and fundamental research.

\begin{acknowledgments}		
For their expertise and competent fabrication of numerous parts we gratefully acknowledge the MPIK engineering design office headed by Frank M\"uller, the MPIK mechanical workshop led by Thorsten Spranz, the MPIK mechanical apprenticeship workshop led by Stefan Flicker and Florian S\"aubert, and the MPIK electronics apprenticeship workshop as well as electronics workshop with Jochen Stephan and Thomas Busch. Likewise, we acknowledge the PTB scientific instrumentation department 5.5 headed by Frank L\"offler, in particular, we thank Stephan Metschke. We also appreciate expeditious technical help by Christian Kaiser. Financial support was provided by the Max-Planck-Gesellschaft, in particular, through IMPRS-PTFS and IMPRS-QD, the Physikalisch-Technische Bundesanstalt, the Bundesministerium f\"ur Bildung und Forschung through project 05K13SJ2, the Deutsche Forschungsgemeinschaft through SFB 1225 ISOQUANT and SCHM2678/5-1, and JSPS Grants-in-Aid for Scientific Research through 24740278 and 15K04707.
\end{acknowledgments}

\end{document}